\documentclass[twocol]{ametsoc}
\journal{jas}
\pdfoutput=1

\bibpunct{(}{)}{;}{a}{}{,}

\usepackage{overpic}
\usepackage{mathrsfs}
\usepackage{caption}
\usepackage{pifont}
\usepackage{comment}

\makeatletter
\def\amsbb{\use@mathgroup \M@U \symAMSb}
\makeatother
\usepackage{bbm}
\usepackage{amsmath,amsfonts}
\makeatletter
\def\amsbb{\use@mathgroup \M@U \symAMSb}
\makeatother
\usepackage[bbgreekl]{mathbbol}
\newcommand\ve[1]{\boldsymbol{#1}}
\newcommand{\ma}[1]{\ensuremath{\amsbb{#1}}}

\newcommand{\gghat}{\ensuremath{\hat{\ve g}}}

\newcommand\nn{\nonumber}

\newcommand{\rd}{{\rm d}}

\newcommand{\eqnlab}[1]{\label{eq:#1}}

\newcommand{\eqnref}[1]{(\ref{eq:#1})}

\newcommand{\Eqnref}[1]{Eq.~(\ref{eq:#1})}
\newcommand{\Figref}[1]{Fig.~\ref{fig:#1}}

\newcommand\etaK{\eta_{\rm K}}
\newcommand\tauK{\tau_{\rm K}}
\newcommand\taus{\tau_{\rm s}}
\newcommand\taup{\tau_{\rm p}}
\newcommand\taud{\tau_{\rm d}}
\newcommand\tauphi{\tau_{\varphi}}
\newcommand{\tr}{{\rm Tr}}
\newcommand{\ku}{{\rm Ku}}
\newcommand{\st}{{\rm St}}
\newcommand{\sv}{{\rm Sv}}
\newcommand{\re}{{\rm Re}}
\newcommand{\St}{{\rm St}}
\newcommand{\rep}{{\re_{p}}}
\newcommand\ReLambda{{\rm Re}_{\lambda}}
\newcommand{\asp}{\beta}

\newcommand\Ao{A_\perp}

\newcommand\Co{C_\perp}
\newcommand\Io{I_\perp}

\newcommand\AgNew{A^{(g)}}
\newcommand\ApNew{A^{(p)}}

\newcommand\dphi{\delta\varphi}

\newcommand\Cp{C_\parallel}
\newcommand\Ip{I_\parallel}

\newcommand\phizerostar{\varphi^\ast}

\def\onedot{$\mathsurround0pt\ldotp$}
\def\cddot{
  \mathbin{\vcenter{\baselineskip.67ex
    \hbox{\onedot}\hbox{\onedot}}%
     }}%

\title{Effect of particle inertia on the alignment of small ice crystals in turbulent clouds}
\authors{K. Gustavsson}
\affiliation{Department of Physics, Gothenburg University, 41296 Gothenburg, Sweden.}
\extraauthor{M. Z. Sheikh}
\extraaffil{\small Univ. Lyon, ENS de Lyon, Univ. Claude Bernard, CNRS, Laboratoire de Physique, F-69342, Lyon, France.}
\extraauthor{A. Naso}
\extraaffil{\small Univ. Lyon, Ecole Centrale de Lyon, Univ. Claude Bernard, CNRS, INSA de Lyon, Laboratoire de M\'ecanique des Fluides et d'Acoustique, F-69134, Ecully, France.}
\extraauthor{A. Pumir}
\extraaffil{\small Univ. Lyon, ENS de Lyon, Univ. Claude Bernard, CNRS, Laboratoire de Physique, F-69342,
Lyon, France.}
\extraauthor{B. Mehlig}
\extraaffil{\small Department of Physics, Gothenburg University, 41296 Gothenburg, Sweden.}

\abstract{Small non-spherical particles settling in a quiescent fluid tend to orient so that their broad side faces down, because this is a stable fixed point of their angular dynamics at small particle Reynolds number. Turbulence randomises the orientations to some extent, and this affects the reflection patterns of polarised light from turbulent clouds containing ice crystals. An overdamped theory predicts that turbulence-induced fluctuations of the orientation are very small when the settling number $\sv$ (a dimensionless measure of the settling speed) is large.  At small $\sv$, by contrast, the overdamped theory predicts that turbulence randomises the orientations. This overdamped theory neglects the effect of particle inertia.  Therefore we consider here how particle inertia affects the orientation of small crystals settling in turbulent air. We find that it  can significantly increase the orientation variance, even when the Stokes number $\st$ (a dimensionless measure of particle inertia) is quite small. We identify  different asymptotic parameter regimes where the tilt-angle variance is proportional to different inverse powers of $\sv$.  We estimate parameter values for ice crystals in turbulent clouds and show that they cover several of the identified regimes. The theory predicts how the degree of alignment depends on particle size, shape and turbulence intensity, and that the strong horizontal alignment of small crystals is only possible when the turbulent energy dissipation is weak, of the order of $1\,$cm$^2$/s$^3$ or less.  }

\begin{document}
\maketitle
\section{Introduction}
As ice crystals settle through turbulent air,
the turbulent velocity gradients tend to randomise their orientations.
However, sometimes the crystals appear to align as they settle, so that they fall with a marked horizontal orientation. The effect can be observed in the form of light patterns above street lights during snow fall \citep{Sassen:80},  \lq{}light pillars\rq{},  caused by specular reflection from the aligned ice-crystal platelets. The width of a light pillar is determined by the degree to which the crystal orientations are randomised.

The alignment of ice crystals in turbulent clouds has been systematically studied
using LIDAR measurements~\citep{Sassen:91,Noel:04,Bre04}. In cirrus
clouds, the fluctuations of the crystal orientation
with respect to the horizontal can be less than a few
degrees~\citep{Sassen:01,Noel:04,Noel:05,Westbrook:2010}.
\cite{Baran:2012} points out that aligned ice crystals affect the way in which clouds
reflect radiation. High-altitude cirrus clouds tend to contain large ice mass. When such clouds cover a non-negligible part of the Earth's atmosphere, ice-crystal alignment could affect its radiation balance, but the magnitude of this effect remains
to be understood.

Hydrodynamic torques due to shape asymmetries or  fluid-inertia  can align the ice crystals.
Rapidly settling particles
experience a locally uniform flow-component (equal to the negative settling velocity).
The resulting fluid-inertia torque tends to orient small fore-aft symmetric and axisymmetric
 particles so that they fall with their broad sides down \citep{Brenner61,Cox65,Kha89,Dab15,Can16,roy2019symmetry}.

Turbulence, on the other hand, may upset the alignment. Early work concluded that turbulence has at most a minor effect on the alignment \citep{Cho81}.
The more recent analysis of \cite{Kle95} was carried out
under the assumption that turbulent torques
act as a white-noise signal on the settling particles.
The resulting diffusion approximation simplifies the analysis,
but it is justified  at very high settling speeds only.

A systematic approach for small particles \citep{Kramel,Men17,Gus19,Anand2020} leads to the prediction of two
very different regimes:
at small settling speeds the orientation is random, while the particles are almost completely aligned at larger settling speeds.
This theory assumes that the dynamics is overdamped, that  particle inertia is negligible, in other words. In this extreme  limit, the particles
move in such a way that the  instantaneous force and torque vanish, and the overdamped
 theory predicts a much stronger alignment at large settling speeds than the theory of \cite{Kle95}.
Is this difference due to the effect of particle inertia, neglected in the overdamped theory, but considered by \cite{Kle95}?
After all, ice crystals are  approximately
1000 times heavier than air, so particle inertia could have a significant effect upon their orientations.

\begin{figure}
\mbox{}\vspace*{1mm}
\begin{center}
\begin{overpic}[width=6cm,clip]{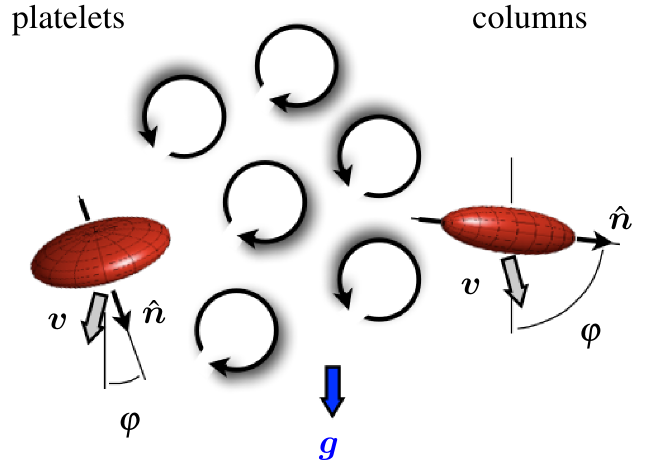}
\end{overpic}
\end{center}
\caption{\label{fig:illustration}
Platelet (left) and column (right) settling in a turbulent flow. The particle
symmetry axis is $\hat{\ve n}$, and the particle velocity is denoted by $\ve v$. Gravity $\ve g = g \hat{\ve g}$ points downwards. The tilt
angle is defined as $\cos\varphi=\pm \hat{\ve n}\cdot\hat{\ve g}$ (see text).
In a quiescent fluid small columns fall with steady-state orientation $\hat{\ve n}\cdot\hat{\ve g}=0$, while
platelets fall with steady-state orientation $\hat{\ve n}\cdot\hat{\ve g}=\pm 1$ (see text).}
\end{figure}

To answer this question, we have investigated the effect of particle inertia upon the alignment of non-spherical particles
settling through a turbulent flow, by analysing a statistical model of the effect.
The main result of our analysis is that particle inertia may lead to
significant  fluctuations of the \lq{}tilt angle\rq{} (Fig.~\ref{fig:illustration}). This effect
results from a coupling between the fluctuations in the translational
dynamics  induced by turbulence, and the angular
degrees of freedom. We find that the overdamped approximation applies only in a small region in the parameter plane.
Even when particle inertia (measured by the \lq{}Stokes number\rq{} $\st$) is weak, it may nevertheless have a substantial
effect on the particle orientation. This is the case when the settling speed is large (large \lq{}settling number\rq{} $\sv$).
Klett's theory fails in this regime because it does not take into account translational particle inertia.

In short, the tilt-angle variance is much larger than previously thought. Particle inertia may increase typical tilt angles by several orders of magnitude compared with the overdamped limit, even at small $\st$, and the theory predicts how typical tilt angles depend upon turbulence intensity, particle size, and shape.
We validate the predictions of our new theory by numerical computations using statistical-model simulations and direct numerical simulations (DNS) of turbulence.

The general conclusion is diametrically opposite to that of
\cite{Cho81}, who concluded that turbulence does not upset the alignment under realistic cloud conditions, in agreement with the prediction of the overdamped theory. Instead, properly taking into account particle inertia, we see that turbulence tends to misalign the orientations of the settling crystals, unless the turbulence level is very weak. This is consistent
with the very strong alignment observed in cirrus clouds~\citep{Noel:04,Noel:05}, because these clouds have a very low
turbulence intensity~\citep{Gultepe:1995}.

Our  theory for ice-crystal platelets  may also explain why only a small fraction of ice crystals appears to align in more turbulent  clouds \citep{Bre04}: the spatially varying conditions must be just right for strong alignment. A caveat, however, is that ie crystals come in a wide variety of shapes, symmetric but also asymmetric, even fractal, and sometimes hollow~\citep{Heymsfield1973}. The fraction of aligned ice crystals, and their average  tilt-angel variance must depend on the distribution of shapes, sizes, and mass-density inhomogeneities.

The remainder of this paper is organised as follows. In Section \ref{sec:background} we give some background. Our model is summarised in Section \ref{sec:model},
including a brief account of the overdamped theory  \citep{Kramel,Men17,Gus19,Anand2020}.
Section \ref{sec:sae} explains our method, an expansion in small tilt angles  \citep{Kle95}.
In Section  \ref{sec:ts} we describe the different physical regimes caused by particle inertia.
Our theoretical results are summarised in Section \ref{sec:results} and discussed in Section \ref{sec:discussion}, which also contains
a detailed comparison with the theory of \cite{Kle95}. Section \ref{sec:conclusions} contains our conclusions.
 A complete summary of our calculations is given in a Supplemental Material.

\section{Background}
\label{sec:background}

Ice crystals come in different shapes \citep{Noel2006}. Frequently observed shapes are columns (rod-like crystals) and platelets (disks) that exhibit discrete rotation symmetry with respect to a symmetry axis $\hat{\ve n}$.
Such platelets correspond to class P1a in the classification of \cite{Magono:66}. 
Commonly such crystals exhibit fore-aft symmetry. This means that particle shape is symmetric
under $\hat{\ve n} \to -\hat{\ve n}$.

A small particle falling in a fluid experiences a mean flow corresponding to its negative settling velocity, plus fluctuations if the fluid is in motion (or is set into motion by the settling particle).
Both mean flow and fluctuating fluid-velocity gradients give rise to torques that affect the orientation of a non-spherical particle. The relative importance of the two torques depends upon
the settling speed, and on the shape of the particle.

The mean flow causes a small axisymmetric particle with fore-aft symmetry and homogeneous mass distribution to align with respect to  the direction of the gravitational acceleration $\ve g$ \citep{Brenner61,Cox65,Kha89,Dab15,Can16}, so that $\hat{\ve n} \perp \ve g$ for columns, and $\hat{\ve n}\parallel\ve g$ for platelets.
The tilt angle is defined as $\cos\varphi=\pm\hat{\ve n}\cdot\hat{\ve g}$ (Fig.~\ref{fig:illustration}).
We denote  its deviations  from the steady-state value by $\delta\varphi$, that
is $\varphi =\delta\varphi$ for platelets and $\varphi = \tfrac{\pi}{2}+\delta\varphi$ for columns.

Several approaches have been proposed to study how turbulence  affects the alignment of settling crystals.
Motivated by the observation that crystal orientation determines the rate
at which crystals are electrically charged, \cite{Cho81} focused on the
vorticity fluctuations in the fluid, neglecting the effect of the turbulent strain, and concluded that turbulence only weakly affects the
crystal orientation. \cite{Kle95} formulated an elegant and more quantitative model describing the effect of
turbulent vorticity and strain \citep{Jef22} upon the orientation of settling crystals.
The model determines how typical tilt angles depend on particle size and turbulent intensity. Klett's theory predicts that the tilt angle
has a narrow distribution. For small particles, its variance decreases as
\begin{equation}
\label{eq:klett}
\langle \delta\varphi^2\rangle \sim \frac{a^2}{\nu}\frac{\mathscr{E}}{W^2}
\end{equation}
as the settling speed $W$ increases. Here $a$ is the particle size,
$\nu$ is the kinematic viscosity of air, and $\mathscr{E}$ is the turbulent dissipation rate per unit mass.
Klett's theory uses an approximate model for the inertial torque  for nearly spherical particles \citep{Cox65}, valid at small
particle Reynolds number in steady flow.
The theory is based on an expansion of the inertial angular dynamics in small $\delta\varphi$. Consistency with Eq.~(\ref{eq:klett}) requires that the settling speed is large so that $\delta\varphi$ remains small.
The  theory also assumes that fluctuations in the settling speed due to translational particle inertia are negligible, and
that the turbulent torques fluctuate very rapidly so that diffusion approximations can be used.

\cite{Gus19} computed the orientation variance in the opposite
limit assuming that the angular dynamics is overdamped and
that the turbulent fluid-velocity gradients experienced by the particle change
slowly compared with the angular dynamics. In this persistent limit
they found for spheroidal columns
\begin{equation}
\label{eq:od1}
\langle \delta\varphi^2\rangle \sim C(\asp)  \frac{\mathscr{E}\nu}{W^4}\,,
\end{equation}
assuming that correlations between $\hat{\ve n}$ and the turbulent fluid velocities are negligible.
The shape parameter $C(\asp)$ in Eq.~(\ref{eq:od1}) is independent of the largest particle dimension, $a$, but it depends on particle shape through the particle aspect ratio $\asp$.
For spherical particles $\asp \to 1$, and in this limit $C(\asp)$ tends to zero. The slender-body limit corresponds to $\asp\to\infty$. In this limit Eq.~(\ref{eq:od1}) was derived by \cite{Kramel} and \cite{Men17}, yielding $C(\asp) \sim \tfrac{32}{375}\log(\asp)^2$.
For platelets ($\beta < 1$), the overdamped approximation works in the same way,
resulting  again  in  Eq.~(\ref{eq:od1}), but with a different prefactor \citep{Anand2020}.

At smaller settling speeds, the settling particles are approximately randomly oriented \citep{Kramel,Gus19}.
In this case, the distribution of $n_g=\hat{\ve n}\cdot\hat{\ve g}$ is uniform, so that one can compute the distribution of tilt angles via a change of variables. The resulting tilt-angle variance is of order unity:
\begin{equation}
\label{eq:random}
\langle \delta\varphi^2\rangle = O(1)\,.
\end{equation}
The transition between Eqs.~(\ref{eq:od1}) and (\ref{eq:random}) is quite sharp. Roughly speaking  the overdamped theory says that the crystals are
either randomly distributed or  well aligned.

\cite{Kramel} measured the orientation variance of nearly neutrally buoyant ramified particles in turbulence,  triads made out of three slender rods. At larger settling speeds the experimental results are roughly consistent with  Eq.~(\ref{eq:od1}), although the data lie somewhat below the theory. Kramel attributed this to the fact that the particles are  larger than the Kolmogorov length and tend to average over small-scale turbulent fluctuations, reducing their effect. \cite{Lop17} measured the orientation distribution of slender columns settling in a two-dimensional steady vortex flow. They showed that the overdamped approximation describes the measured orientation distribution reasonably well. Both experiments were conducted  in water with nearly neutrally buoyant particles, $\rho_{\rm p} /\rho_{\rm f} \approx 1.15$ \citep{Kramel} and $\rho_{\rm p} /\rho_{\rm f} \approx 1.038$ and $1.053$ \citep{Lop17}.

Eq.~(\ref{eq:od1})  predicts a much faster decay of the orientation variance than (\ref{eq:klett}) as the settling speed $W$ increases.
The question is how to reconcile the two estimates.
For ice crystals in air the density ratio is large, $\rho_{\rm p}/\rho_{\rm f} \approx 1000$, so that the overdamped approximation
leading to (\ref{eq:od1}) may break down. Indeed, Eq.~(\ref{eq:od1}) predicts tilt-angle variances that are several orders of magnitude smaller than those observed in turbulent clouds \citep{Bre04}. Simulations of a statistical model for heavy non-spherical particles settling in turbulence
indicate that particle inertia causes Eq.~(\ref{eq:od1}) to fail \citep{Gus19}. Klett's theory takes into account rotational particle inertia, but it also fails to describe the simulation results of \cite{Gus19},
possibly because the theory does not take into account translational particle inertia, which might affect the
alignment indirectly since translation and rotation are coupled.

In summary, it is  likely that rotational and translational particle inertia both have a substantial effect upon the orientation distribution of small crystals settling in a turbulent flow. Yet there is no theory for the effect of particle inertia that is consistent with known limits, and with results of statistical-model simulations. Earlier studies of particles settling in turbulence \citep{Siew14a,Siew14b,Gus17,Jucha2018,Naso2018} included particle inertia, but disregarded the fluid-inertia torque.

\section{Model}
\label{sec:model}
\subsection{Turbulent fluctuations}
\label{sec:tf}
Turbulent flows involve many eddies, covering a wide range of spatial and
temporal scales. The smallest eddies are of the size of the \lq{}Kolmogorov length\rq{}
 $\etaK = (\tfrac{\nu^3}{\mathscr{E}})^{1/4}$.
The fastest time scale associated with the smallest eddies is the Kolmogorov
time, defined as
$ \tauK = [ 2 \langle \tr \, \ma S^2 \rangle ]^{-1/2}$,
 where $\ma S$ is the strain-rate matrix,  the symmetric
part of the fluid-velocity gradient  matrix. Equivalently, one can
simply estimate the Kolmogorov time by $\tauK = ( \tfrac{\nu}{\mathscr{E}})^{1/2}$.

We use a  statistical model \citep{Gus16} for the turbulent fluctuations.
In this model, the fluid-velocity field is represented as
an incompressible Gaussian random function with correlation
length $\ell$, Eulerian correlation time $\tau$ and Lagrangian correlation time  $\tauK=\ell/(\sqrt{5}u_0)$
(\lq{}Kolmogorov time\rq{}). Here $u_0$ is the root mean square turbulent velocity. The model for the turbulent
velocity and gradient correlations is described in the Supplemental Material, see Eq.~(S25).
The correlation length $\ell$ is identified with the Taylor scale $\lambda$  in turbulence.
Since small eddies are swept by larger ones,
 fluid elements advected in turbulent flow decorrelate on the Lagrangian time scale, $\tauK$, not on the Eulerian time scale.
 The statistical model has the same time scale, $\tauK$, if the Kubo number, $\ku=\tau/(\tauK\sqrt{5})$ is large.
  In this limit, the statistical model is therefore expected to
work well \citep{Gus16}, provided that the particles are small enough, with sizes in the dissipative range of turbulence, of the order of $\ell$ and smaller. Otherwise  inertial-range turbulent fluctuations might affect the particle dynamics, and such fluctuations are not taken into account in the statistical model.

Such fluctuations are contained in our model calculations
based on DNS of turbulence, performed using
 the model for the particle dynamics
described in Section \ref{sec:model}.\ref{sec:eom}.
The simulations employ  a  fully dealiased pseudo-spectral code that solves
the Navier-Stokes equation in a  box with periodic boundary
conditions, as described e.g. by \cite{Jucha2018}. The size of the simulation
domain was $L \approx 6.3\,$cm, the viscosity was
$\nu = 0.113 \,$cm$^2$/s, and the turbulent dissipation rate $\mathscr{E} \approx 1\,$cm$^2$s$^{-3}$.
Our simulations were run with a grid of size $128^3$.
This means that they were
well resolved, as can be judged by the value of
$k_{\rm max} \, \etaK \approx 3$, where $k_{\rm max}$ is the largest wave number kept in the Fourier decomposition.
The corresponding Taylor-scale Reynolds number is $\ReLambda \approx 56$.

\subsection{Parameters and dimensionless numbers}
We consider particles with rotational symmetry and fore-aft symmetry.
Commonly observed ice-crystal shapes (columns, platelets) fall into this class \citep{Noel2006}, although more complex shapes
have been reported \citep{Heymsfield2002b}. The dimensions of the settling particle are characterised by the half-length of its symmetry axis, $a_\parallel$, and by the half-length of an orthogonal axis, $a_\perp$. The particle aspect ratio is defined as $\asp= a_\parallel/a_\perp$.
In the following we consider prolate as well as oblate spheroids, $\asp >1$ (columns) and $\asp < 1$ (platelets),
because the hydrodynamic resistance tensors are exactly known for such particles.
This simplification is quite common also in theories regarding other aspects of ice-crystal microphysics, such as crystal growth by vapor deposition~\citep{Chen1994}, and provide a useful approximation to estimate the light reflection properties of crystals~\citep{Yang2013}.
We expect that the theory should work qualitatively for more general columnar and plate-like shapes
\citep{Fri17}. We define the largest particle dimension as
\begin{equation}
\label{eq:amax}
a = {\rm max}\{a_\parallel, a_\perp\}\,,
\end{equation}
and
assume that the particles have uniform mass density~$\varrho_{\rm p}$.
Note, however, that there are ice crystals  in the atmosphere with non-uniform mass densities~\citep{Heymsfield2002}.
\begin{table}[t]
\caption{\label{tab:dimless} Dimensionless parameters.
The time scale $\taup$ is the particle response time, Eq.~(\ref{eq:taup}).
The Kolmogorov scales of the turbulence are denoted
by $\etaK$ and $\tauK$, and $g$ is the magnitude of the gravitational acceleration.
 }
\begin{center}
\begin{tabular}{lll}
\topline
$\asp = a_{\parallel}/a_{\perp}$ & particle aspect ratio  & \\
$a/\etaK$ & particle size &\\
$\varrho_{\rm p}/\varrho_{\rm f}$ &particle-to-fluid density ratio &\\
$\st = \taup/\tauK$ & Stokes number (particle inertia)
&\\
$\sv = g \taup \tauK/\etaK$ & settling number (settling speed)
&\\
$\ell/\etaK$ & turbulent correlation length &\\
\botline
\end{tabular}
\end{center}
\end{table}

In addition to the Reynolds number $\ReLambda$ of the turbulent flow, the problem has at least six additional dimensionless parameters,  summarised in Table~\ref{tab:dimless}. Particle shape is parameterized by its aspect ratio $\asp$. Particle size is parameterised by $a/\etaK$. In the following we assume that this parameter
is small, and we also assume that the particle is much heavier than the fluid
\begin{align}
\label{eq:assumption2}
a/\etaK \ll 1\quad\mbox{and}\quad \varrho_{\rm p}/\varrho_{\rm f} \gg 1\,.
\end{align}
The Stokes number $\st=\taup/\tauK$ is a dimensionless measure of particle inertia,
where
\begin{equation}
\label{eq:taup}
\taup\equiv (2a_\parallel a_\perp\varrho_{\rm p})/(9\nu\varrho_{\rm f})
\end{equation}
 is an estimate of the particle-response time when $ \varrho_{\rm p}/\varrho_{\rm f}\gg 1$.
The settling number $\sv = g \taup \tauK/\etaK$  is a dimensionless measure of the settling speed \citep{Dev12}.
The last parameter is the turbulent correlation length, $\ell/\etaK$.

\subsection{Equations of motion}
\label{sec:eom}
Consider a  small spheroidal particle settling through turbulent air, accelerated
by the gravitational acceleration $\ve g$. The particle is subject
 to a hydrodynamic force $\ve f_{\rm h}$
and to a hydrodynamic torque $\ve \tau_{\rm h}$. Its translational motion
is determined by Newton's second law:
\begin{align}
\label{eq:eomt}
\tfrac{{\rm d}}{{\rm d}t}{\ve x}&= \ve v\,, \quad m\tfrac{{\rm d}}{{\rm d}t}{\ve v} = \ve f_{\rm h} + m \ve g\,.
\end{align}
Here $m$ is the particle mass, $\ve x$ is the spatial position of the particle, and $\ve v$ is its velocity. Particle orientation is defined  by the unit vector
$\hat{\ve n}$ along the symmetry axis of the particle, and its
angular velocity  is denoted by  $\ve \omega$. The angular equations of motion read:
\begin{align}
\label{eq:eomr}
\tfrac{{\rm d}}{{\rm d}t}{\hat{\ve n}}&= \ve \omega \wedge \hat{\ve n}\,,\quad
m\tfrac{\rd}{\rd t}
\big[\ma I(\hat{\ve n}) \ve \omega\big]  = \ve \tau_{\rm h}  \,,
 \end{align}
 where $\ma I(\hat{\ve n})$ is the rotational inertia tensor per unit mass in the lab frame (Supplemental Material).

A major difficulty lies in determining appropriate expressions for the hydrodynamic force and torque.
Here we adopt a simplified model \citep{Kle95,Kramel,Lop17,Men17,Gus19},
adding small inertial corrections due to convective fluid inertia to the standard expressions for  $\ve f_{\rm h}$
and  $\ve \tau_{\rm h}$ in the creeping-flow limit.
In this limit the hydrodynamical force is just Stokes force:
\begin{equation}
\label{eq:stokes}
\ve f^{(0)}_{\rm h} = 6\pi a_\perp\mu \ma A (\hat{\ve n})\big(\ve u-\ve v\big)\,,
\end{equation}
where   $\ve u\equiv\ve u(\ve x,t)$ is  fluid velocity at the particle position $\ve x$,
and $\ma A(\hat{\ve n})$ is a resistance tensor relating $\ve f^{(0)}_{\rm h}$ and the slip velocity $\ve W = \ve v-\ve u$~\citep{Kim:2005}. Its elements depend on  $\asp$ and $\hat{\ve n}$ (Supplemental Material).
 Since they are of order unity for platelets, Eq.~(\ref{eq:stokes}) shows that Eq.~(\ref{eq:taup}) is a natural estimate of the particle response time for platelets of mass $m\propto \varrho_{\rm p} a_\parallel a_\perp^2$.

The hydrodynamic torque in the creeping-flow limit is \citep{Jef22}:
\begin{equation}
\label{eq:jeffery}
\ve \tau^{(0)}_{\rm h} =  6\pi a_\perp\mu \big[
\ma C(\ve\Omega-\ve\omega)
+\ma H\cddot\ma S\big]\,.
\end{equation}
Here $\ve\omega-\ve\Omega$ is the angular slip velocity,
and $\ve \Omega=\tfrac{1}{2} \ve\nabla \wedge \ve u$ is half the fluid
vorticity at the particle position. It is related to the asymmetric part $\ma O$ of the
matrix of fluid-velocity gradients by
the relation $\ma O\ve r = \ve \Omega \wedge \ve r$.  The symmetric part of the matrix of fluid-velocity gradients is denoted by $\ma S$, as mentioned above. The tensors  $\ma C(\hat{\ve n})$ and $\ma H(\hat{\ve n})$ determine
the coupling of the hydrodynamic torque to vorticity and strain~\citep{Kim:2005}.
They depend on the instantaneous particle orientation $\hat{\ve n}$ and  on $\asp$ (Supplemental Material).

Eqs.~(\ref{eq:stokes}) and (\ref{eq:jeffery}) neglect that the particle accelerates the surrounding
fluid as it settles through the flow.
For a particle falling  through a  fluid with a steady settling velocity, the slip velocity $W$ generates fluid accelerations;
it acts as a homogeneous background flow.
To leading
order in the particle Reynolds number
\begin{equation}
\label{eq:rep}
\rep = aW/\nu{\quad (a = {\rm max}\{a_\parallel, a_\perp\})}\,,
\end{equation}
the resulting steady convective-inertia corrections to the force and torque in a quiescent fluid are \citep{Brenner61,Cox65,Kha89,Dab15}:
\begin{subequations}
\label{eq:inertial_corrections}
\begin{align}
 \label{eq:drag_correction}
&\hspace*{-2mm}\ve f^{(1)}_{\rm h} \!=\! -(6\pi a_\perp\mu ) {\scriptstyle \tfrac{3}{16}}\frac{aW}{\nu}  \big[3\ma A\!-\!
\mathbb{1}
(\hat{\ve W}\!\cdot \ma A \hat {\ve W})\big]\ma A \ve W,\\
&\hspace*{-2mm}\ve\tau^{(1)}_{\rm h}= F(\asp){\mu}   \,\frac{a^3W^2}{\nu}\,
 (\hat{\ve n}\cdot {\hat{\ve W}})(\hat{\ve n}\wedge {\hat{\ve W}})\,.
\eqnlab{torque_fluid_inertia}
\end{align}
\end{subequations}
Here $W = |\ve W|$ is the modulus of the slip velocity, and $\hat{\ve W} = \ve W/W$ is its direction, and  $F(\asp)$  is a shape factor computed
by  \cite{Dab15}.
For slender columns, in the limit of $\asp\to\infty$, the shape factor tends to $F(\asp) \sim -5\pi/[3(\log\asp)^2]$. In this
limit Eq.~(\ref{eq:torque_fluid_inertia}) reduces to the slender-body limit derived by  \cite{Kha89}.
For nearly spherical particles the shape factor behaves as
$F(\asp) \sim \mp 811 \pi\varepsilon/560$ for small eccentricity $\varepsilon$, defined by setting $\asp = 1+\varepsilon$
for prolate particles, and $\asp = (1-\varepsilon)^{-1}$ for oblate particles.

For a particle settling through a fluid, one must in principle consider the inertial effect due to gradients of the undisturbed fluid,
parameterised by the shear Reynolds number $\re_s = a^2 s/\nu$, where $s$ is the shear rate
\citep{subramanian2005,einarsson2015a,rosen15}. If we estimate typical turbulent shear rate by $\tauK^{-1}$, we see that the model requires small particles, with particle sizes of the order of $\eta_{\rm K}$ or smaller \citep{Candelier2016}.

We also neglect possible effects of unsteady fluid inertia, a common approximation in the literature, and simply assume that force and torque on the settling particle are given by adding the steady
inertial contributions (\ref{eq:inertial_corrections}) to Stokes force and Jeffery torque \citep{Kle95,Kramel,Lop17,Men17,Gus19}. \cite{Lop17} demonstrated that this model can qualitatively describe the unsteady angular dynamics of rods settling in a vortex flow. The same model was  used first by  \cite{Kle95} to study the angular dynamics of nearly spherical  particles settling in turbulence (we discuss the relation between Klett's and our own theory in Section \ref{sec:discussion}.\ref{sec:klett}). When the slip velocity varies rapidly, the steady model for the inertial torque may fail because the unsteady term in the Navier-Stokes  equations may be equally or more important than the convective terms. We address this limitation of the model in our discussion, Section \ref{sec:discussion}.\ref{sec:limitations}.

We non-dimensionalise Eqs.~(\ref{eq:eomt}) to (\ref{eq:inertial_corrections}) with $\tauK$ and $\etaK$:
$t' = t/\tauK$, $x'= x/\etaK$.
To simplify the notation we drop the primes. The dimensionless equations of motion read:
\begin{subequations}
\label{eq:eom}
\begin{align}
\label{eq:v_p0}
\tfrac{{\rm d}}{{\rm d}t}{\ve x}&=\ve v\,,\quad
\st\,\tfrac{{\rm d}}{{\rm d}t}{\ve v}=-\ma A \ve W +\sv\gghat\,,\\
\label{eq:omega_p0}\tfrac{\rm d}{{\rm d}t}{\hat{\ve n}}&=\ve\omega\wedge\hat{\ve n}\,,\quad
\st\,\tfrac{{\rm d}}{{\rm d}t}{{\ve\omega}}=\st\,\Lambda(\hat{\ve n}\cdot\ve\omega)(\ve\omega\wedge\hat{\ve n})\\
\nonumber &\mbox{}\hspace*{0mm}+
\ma I^{-1}\ma C(\ve\Omega-\ve\omega)
+\ma I^{-1}\ma H\cddot\ma S
+\mathscr{A}'(\hat{\ve n}\cdot\ve W)(\hat{\ve n}\wedge\ve W)\,,
\end{align}
\end{subequations}
with dimensionless parameters $\st$, $\sv$, and $\asp$. The tensors
$\ma I, \ma A, \ma C$, and $\ma H$ are given in the Supplemental Material.
The shape factor $\mathscr{A}'$ is defined as
\begin{equation}
\label{eq:Aprime}
\mathscr{A}'=\frac{5}{6\pi}F(\asp) \frac{\max(\asp,1)^3}{\asp^2+1}\,,
\end{equation}
 and the parameter $\Lambda=\tfrac{\asp^2 - 1}{\asp^2 + 1}$  was defined by  \cite{Bretherton:1962}.
In Eq.~(\ref{eq:eom}) we neglected the inertial contribution (\ref{eq:drag_correction}) to the hydrodynamic force, but kept the contribution (\ref{eq:torque_fluid_inertia}) to the hydrodynamics torque. In the theory and in the statistical-model simulations, the force corrections are not taken into account.
Our numerical simulations with DNS of turbulence
were performed both with and without the correction \eqnref{drag_correction}.

\subsection{Overdamped limit}
\label{sec:od}
\cite{Gus19} analysed the overdamped limit of a prolate spheroid settling
in turbulence  by taking the limit of $\st\to 0$ in Eq.~(\ref{eq:eom}),  as suggested  by \cite{Lop17}. While \cite{Gus19}  considered arbitrary aspect ratios for columns, $\asp >1$, an equivalent approach was pursued by~\cite{Kramel} and by \cite{Men17} in the slender-body limit $\asp\to\infty$.
In the overdamped limit $\st\to 0$, the equations of motion~(\ref{eq:eom})  take the form:
\begin{subequations}
\label{eq:odeom}
\begin{align}
\label{eq:Weq}
\ve W&=\ve W^{(0)}(\hat{\ve n})=\sv\ma A^{-1}(\hat{\ve n})\gghat\,,\\
\eqnlab{omega_overdamped}
 \ve \omega&=\ve \Omega+\Lambda (\hat{\ve n}\wedge \ma S\hat{\ve n})
+{{\mathscr A}\sv^2} (\hat{\ve n}\cdot\hat{\ve g})(\hat{\ve n}\wedge\gghat)\,,\\
\tfrac{\rm d}{{\rm d}t}{\hat{\ve n}} &= \hat{\ve n}\wedge\ve \omega\,.
\eqnlab{nDot_overdamped}
\end{align}
Here $\ve W^{(0)}(\hat{\ve n})$ is the steady slip velocity in the creeping-flow limit, of a spheroid subject
to the gravitational acceleration $\ve g$.
The shape factor $\mathscr{A} $ is given by
\begin{align}
\label{eq:Adef}
\mathscr{A} = \mathscr{A}'{I_\perp}/(A_\parallel A_\perp C_\perp)\,,
\end{align}
\end{subequations}
where $\mathscr{A}'$ was defined in Eq.~(\ref{eq:Aprime}). The remaining coefficients are elements
of the particle-inertia tensor $\ma I$ and the resistance tensors $\ma A$ and $\ma C$  (Supplemental Material).

Eq.~(\ref{eq:omega_overdamped}) illustrates how the fluid-velocity gradients compete with the torque due to convective fluid inertia.
In the absence of flow, the angular dynamics is consistent with  earlier results \citep{Cox65,Kha89,Dab15,Can16}: for prolate particles it has a stable fixed point at $\hat{\ve n}\cdot\hat{\ve g}=0$. This means that rods settle with their symmetry vector orthogonal to the direction of gravity, $\hat{\ve n} \perp \hat{\ve g}$, as mentioned above.
For oblate particles there are two stable fixed points  at $\hat{\ve n}\cdot\hat{\ve g}=\pm 1$, so that disks settle with their symmetry vector parallel with gravity, $\hat{\ve n} \parallel \hat{\ve g}$. In short, the effect of weak convective fluid inertia causes a small spheroid in a quiescent fluid to settle with its broad side first.

Turbulent velocity gradients modify the instantaneous fixed points of the angular dynamics, they change as the particle settles through the flow. The  particle orientation $\hat{\ve n}$ follows the fixed points quite closely if the fluid-velocity gradients change slowly compared to the stability time of the fixed point.
This condition is satisfied for small $\st$ and large $\sv$. At first sight this may seem surprising because
the fluid-velocity gradients vary very rapidly when $\sv$ is large. But note that the stability time is even smaller,
because of the $\sv^2$-factor in \Eqnref{omega_overdamped}).
In this limit the variance of the tilt angle $\delta\varphi$ follows from the statistics of the fluid-velocity gradients.
For columns \cite{Gus19} found:
\begin{equation}
\label{eq:largesvod}
\langle \delta\varphi^2\rangle = \frac{\langle O_{12}^2\rangle+{\Lambda^2}\langle S_{12}^2\rangle}{(\mathscr{A}\sv^2)^2}\propto \frac{\mathscr{E}\nu}{W^4}\,.
\end{equation}
At large settling numbers one may neglect preferential sampling to obtain $\langle O_{12}^2\rangle=\tfrac{5}{3}\langle S_{12}^2\rangle=\tfrac{1}{12}$ for isotropic homogeneous flows. Using $W\sim\ve W^{(0)}(\gghat)=\sv/\Ao$ determines the shape parameter in Eq.~(\ref{eq:od1}), namely $C(\asp) = \tfrac{5+3{\Lambda^2}}{60}\mathscr{A}^{-2}{\Ao^{-4}}$ for columns.
As $\asp\to\infty$ we obtain  $C(\asp)\sim \tfrac{32}{375}\log(\asp)^2$, so that Eq.~(\ref{eq:largesvod}) is consistent with
the slender-body limit derived earlier by \cite{Kramel}.
In homogeneous isotropic flows, $C(\asp)$  is twice as large for platelets, compared with columns \citep{Anand2020}.

\section{Small-angle expansion}
\label{sec:sae}
When $\sv$ is large,
we expect the inertial torque to dominate the angular dynamics, leading to strong alignment of the settling crystals. In this limit the tilt-angle distribution  is sharply peaked around  $\phizerostar = \tfrac{\pi}{2}$
for columns, and around $\phizerostar=0$ for platelets.
In this case, it is sufficient to consider small deviations $\delta\varphi = \varphi-\phizerostar$  from the steady-state angle, and to expand
Eqs.~(\ref{eq:eom}) for $|\delta\varphi | \ll 1$ as first suggested by \cite{Kle95}.
In the following we restrict the range of $\varphi$ to $0\le\varphi\le\pi$ for columns, and to $-\pi/2\le\varphi<\pi/2$ for platelets. Negative values of $\varphi$ correspond to $\hat{\ve n}\cdot\hat{\ve g} < 0$.
\begin{figure}[t]
\mbox{}\vspace*{1mm}
\begin{center}
\begin{overpic}[width=4.25cm,clip,angle=0]{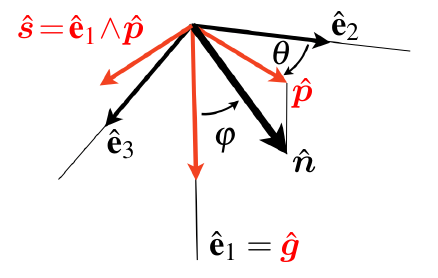}
\end{overpic}
\mbox{}\vspace*{-3mm}
\end{center}
\caption{\label{fig:coordinates}  Coordinate system for angular dynamics: direction of gravitational acceleration $\hat{\ve g} = \hat{\bf e}_1$, projection $\hat{\ve p}$ of $\hat{\ve n}$ onto the plane perpendicular to gravity, and $\hat{\ve s} = \hat{\bf e}_1\wedge \hat{\ve p}$.
 }
\end{figure}

A convenient coordinate system for the analysis is illustrated
in Fig.~\ref{fig:coordinates}. Namely, we take as coordinate axes
gravity ($\gghat={\ve g}/|{\ve g}|$), the projection $\hat{\ve p}$  of $ \hat{\ve n}$ onto the plane perpendicular to gravity (so that $\hat{\ve n} = \hat{\ve g} \cos\varphi + \hat{\ve p} \sin\varphi$ for $\varphi>0$), and $\hat{\ve s} = \hat{\ve g}\wedge \hat{\ve p}$. In this coordinate system, the
components of $\hat{\ve n}$ are $n_g, n_p$, and $n_s=0$. We denote the corresponding components of other vectors and tensors using similar subscripts.
We assume that the gravitational acceleration  points in the $\hat{\bf e}_1$-direction.
The components of the particle-symmetry axis $\hat{\ve n}$ read
\begin{equation}
\label{eq:npa}
{\hat{\ve n}} = {\rm sgn}(\varphi)\begin{bmatrix} \cos\varphi\\\sin\varphi \cos\theta\\\sin\varphi\sin\theta\end{bmatrix}.
\end{equation}
The orientation vector $\hat{\ve n}$
is determined by two angles, the tilt angle $\varphi$, and the angle $\theta$ describing the orientation of the particle-symmetry vector $\hat{\ve n}$ in the plane orthogonal to gravity.
The factor sgn$(\varphi)$ is not strictly necessary, but it is convenient because it allows us to use Eq.~(\ref{eq:npa}) to parameterise
$\hat{\ve n}$ for both columns and for platelets.

We project the angular dynamics \eqref{eq:omega_p0} onto the basis vectors $\hat{\ve g}$, $\hat{\ve p} $, and $\hat{\ve s} $, and expand
to linear order in $\delta\varphi$. For platelets this gives:
\begin{subequations}
\eqnlab{omega_spherical_smalldphi}
\begin{align}
\begin{split}
\tfrac{{\rm d}}{{\rm d}t}\dphi&=\omega_s\,,\quad \tfrac{{\rm d}}{{\rm d}t}\theta=-{\omega_p}/{\dphi}\,,\\
\tfrac{{\rm d}}{{\rm d}t}\omega_s &=\tfrac{\Co}{\Io\st}(-\omega_s + Y_{gp} - Y_{gg}\dphi) + {\omega_p^2}/{\dphi}\,,\\
\tfrac{{\rm d}}{{\rm d}t}\omega_p &=\tfrac{\Co}{\Io\st}(-\omega_p - Y_{gs}) - {\omega_p\omega_s}/{\dphi}\,,\\
\tfrac{{\rm d}}{{\rm d}t}\omega_g &= \tfrac{\Cp}{\Ip\st}(-\omega_g +\Omega_g) + \tfrac{\dphi}{\st}\tfrac{\Co}{\Io}Y_{gs}\,.
\end{split}
\eqnlab{omega_spherical_smalldphi_platelets}
\intertext{For columns we obtain:}
\begin{split}
\tfrac{{\rm d}}{{\rm d}t}\dphi&=\omega_s\,,\quad\tfrac{{\rm d}}{{\rm d}t}\theta=\omega_g\,,\\
\tfrac{{\rm d}}{{\rm d}t}\omega_s &=\tfrac{\Co}{\Io\st}(-\omega_s + Y_{gp} - Y_{gg}\dphi)\,,\\
\tfrac{{\rm d}}{{\rm d}t}\omega_p &=\tfrac{\Cp}{\Ip\st}(-\omega_p + \Omega_p)\,,\\
\tfrac{{\rm d}}{{\rm d}t}\omega_g &= \tfrac{\Co}{\Io\st}(-\omega_g - Y_{sp}+ \dphi Y_{gs})\,.
\end{split}
\eqnlab{omega_spherical_smalldphi_columns}
\end{align}
\end{subequations}
In this small-$\delta\varphi$ expansion we neglected all terms of second and higher order in $\delta\varphi$. Amongst the terms linear in $\delta\varphi$ we
kept only those proportional to $W_g$, in keeping with our assumption that $\sv$ is large.
Amongst the terms quadratic in the angular velocity we kept only those terms that are multiplied by $\dphi^{-1}$, the other quadratic terms are negligible unless $\st$ is  large.
Finally, we simplified the $\theta$-dynamics for platelets, \Eqnref{omega_spherical_smalldphi_platelets}, neglecting a term proportional to $\omega_g$
which is negligible compared to  $-\omega_p/\dphi$ when $\delta\varphi$ is small.

Eqs.~\eqnref{omega_spherical_smalldphi} are driven by the matrix $\ma Y$, representing fluctuations of the fluid-velocity gradients ($\ma O$ and $\ma S$), and of the slip velocity $\ve W$.
In the Cartesian basis, the elements of $\ma Y$ read:
\begin{align}
Y_{ij} = |{\mathscr A}|\AgNew\ApNew W_i W_j -O_{ij} -|\Lambda|S_{ij}\,.
\eqnlab{YDef}
\end{align}
We see that $\ma Y$  represents two distinct origins of stochasticity. The first term on the r.h.s. of \Eqnref{YDef} stems from the fluctuations of the slip velocity $\ve W$.
The two remaining terms model the effect of the  turbulent fluid-velocity gradients, through the elements $O_{ij}$ and $S_{ij}$
of $\ma O$ and $\ma S$.

\begin{table}[t]
\centering
\captionof{table}{\label{tab:TimeScales} Time scales for Eq.~\eqnref{eom} at large $\sv$ (see text). }
\begin{tabular}{ll}
\topline
Time scale & parameter dependence\cr
\hline
\\[-3.8mm]
fluid-velocity gradients & $\tauK=1$\cr
settling & $\taus={\AgNew\ell}/{\sv}$\cr
fluid-inertia torque & $\tauphi= {1}/{(|\mathscr A|\sv^2)}$\cr
damping & $\taud^{({\rm tr})}={\st}/{\AgNew}$ \hfill(translation) \cr
               & $\taud^{({\rm rot})}={\Io \st}/{\Co}$ \hfill(rotation)\cr
\\[-2.8mm]\botline
\end{tabular}
\end{table}

\section{Analysis of time scales and physical regimes}
\label{sec:ts}
\Eqnref{omega_spherical_smalldphi} has four relevant time scales.
First, the Kolmogorov time $\tauK$ (equal to unity in our dimensionless units) determines the magnitude of the fluid-velocity gradients.
When the settling number $\sv$ is small, $\tauK$ also determines the order of magnitude of the Lagrangian correlation time of tracer particles, of the same
order as $\tauK$, but usually somewhat larger.

Second, when $\sv$ is large, the fluid velocity and the gradients seen by the settling particle decorrelate on  the settling time scale $\taus$.
\cite{Gus19} and \cite{Kramel} estimated $\taus$ as the time it takes to fall one flow-correlation length $\ell$ with settling velocity \Eqnref{Weq} in the steady-state orientation in a quiescent fluid:
\begin{align}
\taus=\AgNew\ell/\sv\,.
\eqnlab{taus}
\end{align}

Third, $\tauphi$ describes the time scale of the  fluid-inertia torque.
In the overdamped limit the angular dynamics is determined by \Eqnref{omega_overdamped}.
Because the fluid-velocity gradients are of order $\sim\tauK^{-1}=1$, the fluid-inertia torque dominates when $|\mathscr A|\sv^2\gg 1$.
\cite{Gus19} used
\begin{align}
\tauphi\equiv 1/(|\mathscr A|\sv^2)\,,
\end{align}
 the time  it takes the overdamped angular dynamics to approach its steady state in a frozen flow. We expect that this remains a reasonable estimate of $\tauphi$ even outside
 the overdamped limit, provided that $\st$ is not too large. This time scale is related to $\tau_{\rm sed}$ considered by \cite{Kramel}, averaged
 over orientations.
 \begin{figure}[t]
\centering
\begin{overpic}[width=7cm,clip]{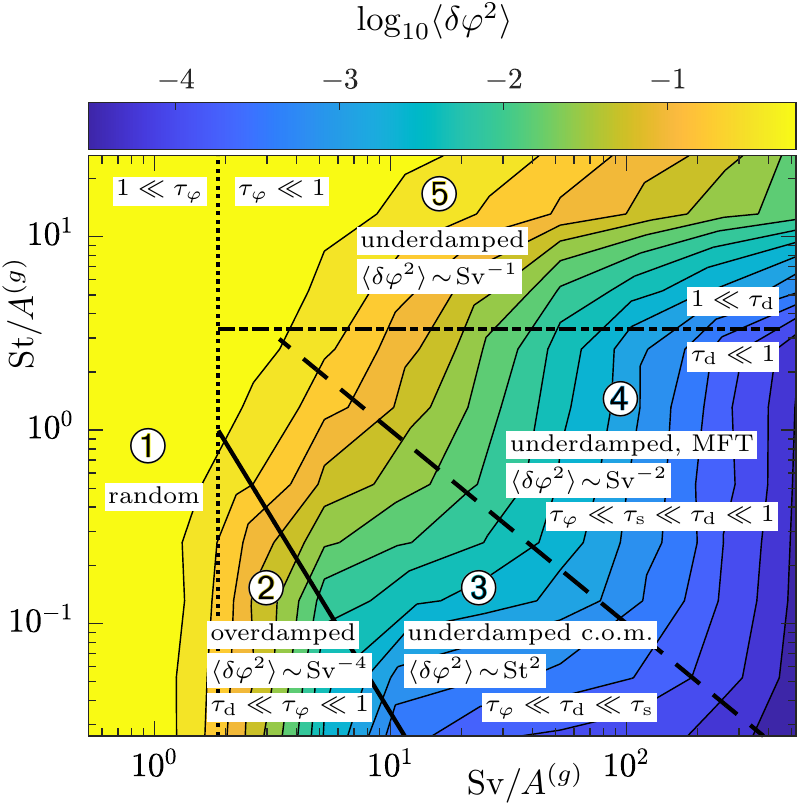}
\end{overpic}
\caption{\label{fig:PhaseDiagram}
Phase diagram of  asymptotic regimes for the tilt-angle variance $\langle\dphi^2\rangle$ in the statistical model, together with results of numerical statistical-model simulations of Eqs.~\eqnref{eom} for platelets with $\asp=0.1$, $\ell=10$, and $\ku=10$ (colour coded, see legend).
The conditions separating the different regimes are discussed in the text:
$\tauphi=1$ (dotted line),
$\tauphi=\taud^{({\rm tr})}$ (solid line),
$\taud^{({\rm tr})}=\taus$ (dashed line), and
$\taud^{({\rm rot})}=1$ (dash-dotted line). \lq{}MFT\rq{} stands for mean-field theory (Section \ref{sec:ts}.\ref{sec:mft}).
}
\end{figure}

Finally, the damping time scale describes the time scale of inertial effects in \Eqnref{omega_spherical_smalldphi}. In dimensionless units this time scale equals $\st$, up to a prefactor determined by the shape coefficients in \Eqnref{omega_spherical_smalldphi}:
\begin{align}
 \taud=\left \{\begin{array}{ll} \st/\AgNew & {\rm translation}\,,\\
 \Io \St/\Co & {\rm rotation}\,.\end{array}\right .
 \end{align}
As long as $\asp$ is not too large, the coefficients $\AgNew$ and $\Co/\Io$ are of the same order for spheroids,
so that $\taud^{({\rm tr})}$ and $\taud^{({\rm rot})}$ are of the same order. Where the quantitative difference matters we distinguish these time scales, otherwise we just write $\taud$.
We remark that even though $\taud^{({\rm tr})}$ and $\taud^{({\rm rot})}$ are of the same order, they may in certain cases affect the rotational and translational dynamics differently. This is discussed below.

The dependence of these time scales upon the dimensionless parameters $\sv$, $\st$,  $\asp$, and $\ell$ is summarised in Table~\ref{tab:TimeScales}.
Comparing the time scales, we identify a number of asymptotic regimes of the angular dynamics \eqnref{omega_spherical_smalldphi} with qualitatively different physical behaviours.
The different regimes are summarised in \Figref{PhaseDiagram}.
The Figure also shows the tilt-angle variance obtained from numerical statistical-model simulations of Eqs.~\eqnref{eom}
 (colour-coded contour plot), as well as the asymptotic statistical-model predictions derived in the following Sections.
We see that the variance ranges over four orders of magnitude for the parameter ranges considered, and that there are five different asymptotic regimes with different mechanisms
at work, leading to distinct scaling predictions for the variance.

\subsection{Random orientation (regime \raisebox{-0.2mm}{\ding{192}})}
When $\sv$ is small so that $\tauphi \gg1$, the crystals are essentially randomly oriented as described in Section \ref{sec:background}. In this regime
the particle orientations are randomised by the turbulent fluid-velocity gradients. The symmetry-breaking  torque due to settling does not matter, so that the tilt angles are randomly distributed with $\langle \delta\varphi^2\rangle \sim O(1)$, Eq.~(\ref{eq:random}).

\subsection{Overdamped dynamics (regime \raisebox{-0.2mm}{\ding{193}})}
\label{sec:ts_overdamped}
When $\tauphi \ll1$ and in addition $\taud \ll \min \{\tauphi,\taus\}$ then both angular and translational dynamics are overdamped.
The persistent limit analysed by \cite{Gus19}  corresponds to $\tauphi\ll\taus$
($\taus\ll\tauphi$ can only occur for nearly spherical particles, see Section \ref{sec:discussion}.\ref{sec:klett}).
When $\tauphi$ is much smaller than $\taus$, the fluid-velocity gradients
remain constant during the time it takes for the tilt angle to adjust to its fixed point. The tilt-angle variance is determined by a balance between the turbulent fluid-velocity gradients and the inertial torque, and the variance is given by Eq.~\eqnref{largesvod} tor columns.

 \subsection{Underdamped centre-of-mass~dynamics (regime \raisebox{-0.2mm}{\ding{194}})}
\label{sec:ts_underdamped_translational}
The asymptotic regime  \raisebox{-0.2mm}{\ding{194}} is delineated by the inequalities
$\tauphi\ll\taud\ll\taus$ and $\taud\ll 1$. Since $\taus\ll{{\rm min}(1,\taus)}$, the angular dynamics is overdamped.
But  since $\taud\gg\tauphi$, the overdamped approximation  \eqnref{Weq} for the slip velocity does not apply,  because the centre-of-mass (c.o.m.)~dynamics does not have time to adjust to the rapid changes in $\dphi$. In this regime the tilt-angle variance is determined by the fluctuations of the underdamped c.o.m.~dynamics, and therefore the variance
depends only weakly on $\sv$, but  strongly on $\st$.
The Jeffery torque \eqnref{jeffery}  plays no role in this regime.

\subsection{Underdamped angular and c.o.m.~dynamics, mean-field theory (regime~\raisebox{-0.2mm}{\ding{195}})}
\label{sec:mft}
Passing from regime  \raisebox{-0.2mm}{\ding{194}} to  \raisebox{-0.2mm}{\ding{195}} in \Figref{PhaseDiagram}, $\taud$ becomes larger than $\taus$.  As a consequence,  both c.o.m. and  angular dynamics are underdamped. In this case the fluid velocity  seen by the particle fluctuates more rapidly than the damping time scale.
Within  a mean-field theory (MFT), see Section II.C in the Supplemental Material, we find that the tilt--angle variance decays as $\sv^{-2}$, just like Eq.~(\ref{eq:klett}). But the prefactor is different
from Klett's prediction (the Jeffery torque \eqnref{jeffery} does not matter in this regime).

\subsection{Underdamped angular and c.o.m.~dynamics (regime \raisebox{-0.2mm}{\ding{196}})}
Regime  \raisebox{-0.2mm}{\ding{196}}~in \Figref{PhaseDiagram} corresponds to $\tauphi\gg 1$ and $\taud\gg~\!\!1$.
So the fluid-inertia torque dominates in this regime, and both c.o.m. and  angular dynamics are underdamped.
  When in addition $\taud\gg\taus$, then the variance of the tilt angle decays as $\sv^{-1}$. We note that this asymptote
  is only reached for the largest $\st$ in Fig.~\ref{fig:PhaseDiagram}.
  The opposite case, $\taud\ll\taus$, is very difficult to realise when $\taud\gg 1$ and $\tauphi\gg 1$.

In summary, the  asymptotic regimes in Fig.~\ref{fig:PhaseDiagram} exhibit different power-law dependencies of the tilt-angle
variance upon the settling number $\sv$. The different power-law scalings are visible as evenly spaced vertical or horizontal level curves in Fig.~3 (see also Fig. S1 in the Supplemental Material which shows the power laws more clearly). Since $\sv\propto \taup\propto a^2$, these statistical-model predictions translate into
different power laws as a function of particle size. The overdamped regime~\raisebox{-0.2mm}{\ding{193}} has the strongest
dependence on particle size, $\langle \delta\varphi^2\rangle \propto a^{-8}$.
However, Fig.~\ref{fig:PhaseDiagram} shows that regime~\raisebox{-0.2mm}{\ding{193}} is quite narrow,
and in regimes~\raisebox{-0.2mm}{\ding{194}} and \raisebox{-0.2mm}{\ding{195}} the variance decays more slowly with increasing particle size.  The  same conclusion holds for the transition from~\raisebox{-0.2mm}{\ding{192}} to~\raisebox{-0.2mm}{\ding{196}}.

\section{Results}
\label{sec:results}
To determine
 the tilt-angle variance in regimes  \raisebox{-0.2mm}{\ding{194}},  \raisebox{-0.2mm}{\ding{195}}, and  \raisebox{-0.2mm}{\ding{196}},
 we solved the  angular dynamics \eqnref{omega_spherical_smalldphi} together with that of  $Y_{ij}$
 [Eq.~\eqnref{YDef}]. A brief yet complete account of our calculations is given in the Supplemental Material. The result is:
\begin{align}
\eqnlab{phiSqrPlateletsUnevaluated}
&\langle\dphi^2\rangle=
f_\Lambda\Big\{\frac{\AgNew{}^2}{\sv^2}C_u(0)
+\frac{\AgNew{}^2}{\ApNew{}^2|\mathscr A|^2\sv^4}C_B(0)
\\&
+\frac{\AgNew}{|\mathscr A|^2 \st \sv^4}\int_0^\infty {\rm d}t e^{-\AgNew t/\st}\Big[\Big(1-\frac{\AgNew{}^2}{\ApNew{}^2}\Big)C_B(t)
\nn\\&
\hspace{1.2cm}+2\AgNew |\mathscr A| \sv C_X(t)
-\AgNew{}^2 |\mathscr A|^2 \sv^2 C_u(t)\Big]\Big\}\,.
\nn
\end{align}
Here $f_\Lambda=2$ for $\Lambda < 0$ (platelets) and $f_\Lambda=1$ for  $\Lambda > 0$ (columns).
Eq.~\eqnref{phiSqrPlateletsUnevaluated}
 is expressed in terms of correlation functions of fluid velocities and fluid-velocity gradients  evaluated along settling trajectories, $C_B(t)=\langle O_{12}(t)O_{12}(0)+2|\Lambda|O_{12}(t)S_{12}(0)+\Lambda^2S_{12}(t)S_{12}(0)\rangle$, $C_u(t)=\langle u_{2}(t)u_{2}(0)\rangle$ and $C_X(t)=\langle u_2(t)[O_{12}(0)+|\Lambda|S_{12}]\rangle$.
 For the statistical model, the correlation functions are given in the Supplemental Material.
 We remark that the average  of the tilt angle and all higher odd-order moments must vanish, because
positive and negative values of $\mbox{sgn}({\delta}\varphi)$ are equally likely,

Eq.~\eqnref{phiSqrPlateletsUnevaluated} shows how translational particle inertia affects the tilt-angle variance.
The flow-velocity correlations in \Eqnref{phiSqrPlateletsUnevaluated} can be traced back to the effect of the fluctuating settling velocity due to particle inertia
[first term on the r.h.s. of Eq.~(\ref{eq:YDef})].
The gradient correlations in \Eqnref{phiSqrPlateletsUnevaluated} stem from the Jeffery torque \eqnref{jeffery}, corresponding to the other two  terms on the r.h.s. of Eq.~(\ref{eq:YDef}).

\Eqnref{phiSqrPlateletsUnevaluated} simplifies to (\ref{eq:largesvod})
when translational inertia is negligible, in regime \raisebox{-0.2mm}{\ding{193}} in \Figref{PhaseDiagram}.
This can be seen by taking the limit $\st/\AgNew\to 0$ in \Eqnref{phiSqrPlateletsUnevaluated}. Using $ \tfrac{\AgNew}{\st}e^{-t\AgNew/\st}\sim  2\delta(t)$ gives
\begin{subequations}
\label{eq:limits}
\begin{align}
\label{eq:odlimit}
 \langle \delta\varphi^2\rangle \sim f_\Lambda  \frac{\langle O_{12}^2\rangle+{\Lambda^2}\langle S_{12}^2\rangle}{(\mathscr{A}\sv^2)^2}\,,
 \end{align}
for columns the same as \Eqnref{largesvod}. For platelets the variance is twice as large, consistent with the result of \cite{Anand2020}. This difference in the prefactor between columns and platelets is a direct consequence of the different dynamics of $\hat{\ve p}$.

In regimes  \raisebox{-0.2mm}{\ding{194}} and  \raisebox{-0.2mm}{\ding{195}},  fluctuations of the translational slip velocity dominate.
This follows from taking the limit $\tauphi\to 0$ in \Eqnref{phiSqrPlateletsUnevaluated}, where contributions from the fluid-velocity gradients disappear.   We distinguish two cases.

First, in regime  \raisebox{-0.2mm}{\ding{194}} ,  we use $\tauphi\ll 1$ to simplify \Eqnref{phiSqrPlateletsUnevaluated}. Integration by parts, rescaling the integration variable with $\taud$,
and using  $\taud\ll\taus\ll1$ to expand the  correlation functions gives:
\begin{align}
\begin{split}
\langle\dphi^2\rangle
&\sim  f_\Lambda \frac{\st^2}{\AgNew{}^2}\langle {A_{21}^2}\rangle
\end{split}\,.
\eqnlab{phiSqrPlateletsPlateau}
\end{align}
\Eqnref{phiSqrPlateletsPlateau} shows that the variance of the tilt angle forms an $\sv$-independent plateau in regime  \raisebox{-0.2mm}{\ding{194}}. In this regime the angular dynamics is overdamped. The $\st$-dependence is caused by
fluctuations in the translational slip velocity, due to particle inertia.

Second, regime  \raisebox{-0.2mm}{\ding{195}} corresponds to $\taus\to 0$ at finite $\taud=\st/\AgNew\ll 1$.
Now the rotational dynamics is underdamped. Nevertheless, 
Equation \Eqnref{phiSqrPlateletsUnevaluated}  continues to hold, as demonstrated by the
mean-field analysis described in Section II.C in the Supplemental Material.

Using the statistical-model correlation functions given in
the Supplemental Material, we find:
\begin{align}
\langle\dphi^2\rangle&\sim f_\Lambda \frac{\AgNew{}^2}{\sv^2}\langle u_2^2\rangle\,.
\eqnlab{phiSqrPlateletsLargeSv}
\end{align}
\end{subequations}
In dimensionless units, for homogeneous isotropic turbulent flows, $\langle u_2^2 \rangle \approx {\rm Re}_\lambda \approx \ell^2/\sqrt{15}$. Accordingly, \Eqnref{phiSqrPlateletsLargeSv} predicts that the tilt-angle variance is proportional to $\sv^{-2}$  in regime  \raisebox{-0.2mm}{\ding{195}}.
Finally, we can evaluate \Eqnref{phiSqrPlateletsUnevaluated} in closed form for the statistical model, exhibiting how the variance depends
on the dimensionless parameters $\st$, $\sv$, and $\asp$ (details in the Supplemental Material).

Our time-scale analysis in Section \ref{sec:ts} led to the phase diagram  \Figref{PhaseDiagram}, describing the asymptotic behaviours of the tilt-angle variance.
We obtain the same asymptotic  boundaries by comparing the corresponding
limits of our theory.
For example, since $\langle u_2^2\rangle\sim\ell^2\langle A_{21}^2\rangle$,
Eqs. \eqnref{phiSqrPlateletsPlateau} and \eqnref{phiSqrPlateletsLargeSv} are equal when $\taus\sim\taud$, the boundary between regimes  \raisebox{-0.2mm}{\ding{194}} and \raisebox{-0.2mm}{\ding{195}}.
Similarly, Eqs. \eqnref{phiSqrPlateletsPlateau} and \eqnref{largesvod} are equal when $\taud\sim\tauphi$, i.e. the boundary between regimes  \raisebox{-0.2mm}{\ding{193}} and \raisebox{-0.2mm}{\ding{194}}.

To obtain an asymptotic law in regime \raisebox{-0.2mm}{\ding{195}} we took the limit $\sv\to\infty$. It is important to note that
the steady approximation for the convective-inertia torque breaks down
when $\ve W$ varies too rapidly (too large $\sv$ gives too small $\taus$). The model requires that  $\taus$ is much larger
than the viscous time, $a^2/\nu$. We discuss this constraint further in Section \ref{sec:discussion}.\ref{sec:limitations}.

\Eqnref{phiSqrPlateletsUnevaluated} does not apply in regime \raisebox{-0.2mm}{\ding{196}} where
both c.o.m. and angular dynamics are underdamped. The settling velocity is large ($\taus$ is small).
When $\taus$ is the smallest time scale we  approximate the $\dphi$-dynamics  as Langevin equations.
Solving the corresponding Fokker-Planck equation for the moments of $\dphi$ we find
in regime \raisebox{-0.2mm}{\ding{196}} for the statistical model:
\begin{align}
\label{eq:5}
 \langle\delta\varphi^2\rangle =f_\Lambda \frac{\sqrt{2\pi}}{60}|\mathscr{A}|\AgNew\ApNew\ell^3 \frac{\AgNew}{ \sv}
 \end{align}
 (details in the Supplemental Material).
 The same caveat as for regime \raisebox{-0.2mm}{\ding{195}} applies:  the settling time $\taus$ must be larger than the viscous time $a^2/\nu$.

\begin{figure}
\centering
\begin{overpic}[width=7.cm,clip]{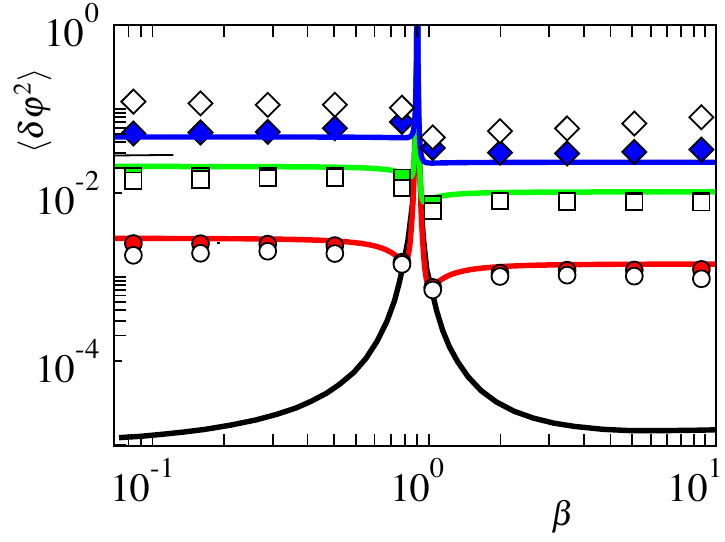}
\end{overpic}
\caption{\label{fig:DNSvariance} Tilt-angle variance
as a function of particle aspect ratio $\asp$  keeping $\st/\AgNew$ and $\sv/\AgNew$ constant.
Results obtained using DNS of turbulence:
empty symbols are with the inertial drag correction (\ref{eq:drag_correction}),  filled symbols without this correction.
The overdamped  approximation (\ref{eq:largesvod}) is shown as a black solid line.
Also shown is the theoretical prediction \eqnref{phiSqrPlateletsUnevaluated} for regimes
\raisebox{-0.2mm}{\ding{193}} to \raisebox{-0.2mm}{\ding{195}} for $\ell=14.7$, coloured lines. Other parameters: $\sv/\AgNew= 22$ and $\st/\AgNew=0.11$ (red,$\circ$), $0.45$ (green,{\scriptsize$\Box$}), and $2.2$ (blue,$\diamond$).
 }
\end{figure}

Fig.~\ref{fig:DNSvariance} shows how the tilt-angle variance depends on the particle aspect ratio, keeping
 $\st/\AgNew$ and $\sv/\AgNew$ constant. The theoretical prediction \eqnref{phiSqrPlateletsUnevaluated} for regimes
\raisebox{-0.2mm}{\ding{193}} to \raisebox{-0.2mm}{\ding{195}}
is shown for three different Stokes numbers (coloured solid lines).
The overdamped approximation (\ref{eq:largesvod}) is plotted as a black solid line.
We see that it is accurate only in
regime \raisebox{-0.2mm}{\ding{193}},
 for $\asp$ approximately between $0.8$ and $1.2$
when $\st/\AgNew=0.11$. For larger Stokes numbers this range is even narrower. Outside  regime \raisebox{-0.2mm}{\ding{193}}, particle inertia matters. We see that particle inertia increases the  tilt-angle variance by a large factor compared to the overdamped approximation,  by several orders of magnitude for slender columns and thin disks. Also, the tilt-angle variance is independent of $\asp$, unless $\asp$ is  close to unity.
This follows from the fact that amongst all time scales in Table~\ref{tab:TimeScales}, only
$\tauphi$ exhibits a significant $\asp$ dependence at constant $\sv/\AgNew$ and $\st/\AgNew$. Since
$\tauphi$ matters only in regime \raisebox{-0.2mm}{\ding{193}},
it follows that the tilt-angle variance does not depend on $\beta$
in regimes \raisebox{-0.2mm}{\ding{194}} and \raisebox{-0.2mm}{\ding{195}}, as long as $\sv/\AgNew$ and $\st/\AgNew$
are kept constant.

Also shown are results of numerical simulations of Eqs.~(\ref{eq:eomt})  to \eqnref{inertial_corrections}
 using DNS of turbulence. To maintain
 $\taup/\AgNew$ constant, we adjusted the particle size as we changed $\asp$.
 We performed DNS with the inertial correction \eqnref{drag_correction} to the translational dynamics (empty symbols) and without (filled symbols). The Reynolds number was ${\rm Re}_\lambda\approx 56$. For the comparison with the theory we identified $\ell$ with the Taylor scale  and used, in dimensional variables,  ${\lambda}/{\etaK}=15^{1/4} \sqrt{{\rm Re}_\lambda}\approx 14.7$.

Figure~\ref{fig:DNSvariance}  demonstrates that our theory \eqnref{phiSqrPlateletsUnevaluated} describes the DNS results very well,
without any fitting parameter.
For the smaller Stokes numbers [$\st/\AgNew= 0.11$ (circles) and $0.45$ (squares)],
 the  inertial correction  \eqnref{drag_correction} to the translational dynamics is quite small, except at
very small and very large values of $\asp$ -- where $\rep$ is largest.
A simple order-of-magnitude estimate  explains that the Oseen correction  \eqnref{drag_correction}
has only a small effect on  $\langle ( \delta\varphi)^2\rangle$  for the parameters considered here \citep{Gus19}:
 the inertial correction  is smaller than the Stokes force \eqnref{stokes}
by the small  factor $a/\etaK$.

Looking in more detail, we infer from
Fig.~\ref{fig:DNSvariance}  that the Oseen correction appears to decrease the tilt-angle variance somewhat
in regimes  \raisebox{-0.2mm}{\ding{193}},  \raisebox{-0.2mm}{\ding{194}}, and  \raisebox{-0.2mm}{\ding{195}}.
This is explained by the fact that the Oseen correction increases the translational drag and therefore reduces the slip-velocity fluctuations.  Yet the difference remains small for the parameters in Fig.~\ref{fig:DNSvariance}, as mentioned above.

The data for the largest Stokes number agrees less well with Eq.~\eqnref{phiSqrPlateletsUnevaluated}. This is expected because the values $\st/\AgNew= 2.2$ and $\sv/\AgNew= 22$ lie near the boundary to regime  \raisebox{-0.2mm}{\ding{196}} where  Eq.~\eqnref{phiSqrPlateletsUnevaluated} begins to fail
(the Stokes number is not  yet large enough for Eq.~(\ref{eq:5}) to work).

We also see  that the inertial correction  \eqnref{drag_correction} to the translational dynamics
makes a substantial  difference in regime  \raisebox{-0.2mm}{\ding{196}}, where the tilt-angle variance is much larger when the drag correction is included. In part this can be attributed to a larger particle Reynolds number, but in regime  \raisebox{-0.2mm}{\ding{196}}  we do not understand the effect of the correction  \eqnref{drag_correction} in detail.

\section{Discussion}
\label{sec:discussion}

\subsection{Comparison with Klett's theory}
\label{sec:klett}
The main assumptions underlying Eq.~(\ref{eq:klett}) are that translational particle inertia is negligible, that the particles are nearly spherical, and that the driving of the angular dynamics is white noise.

The time-scale analysis in Section~\ref{sec:ts} says that 
 translational inertia can only be neglected  in regimes~\raisebox{-0.2mm}{\ding{192}} and~\raisebox{-0.2mm}{\ding{193}}.
  In regime~\raisebox{-0.2mm}{\ding{192}}, the inertial alignment torque is negligble ($\tauphi \gg 1)$.  Therefore this discussion
  concentrates on the case $\tauphi\ll 1$, where  not only translational but also rotational inertia is negligible. 
  
  In Fig.~\ref{fig:PhaseDiagram} we stipulated that $\taus\ll\taud$ in regime~\raisebox{-0.2mm}{\ding{193}}. But for nearly spherical particles, as considered by Klett, one can have $\taud\ll\taus$, so that  $\taud\ll\taus\ll\tauphi\ll 1$. In this limit, $\taud$ is the smallest time scale. Therefore the dynamics is overdamped as in regime \raisebox{-0.2mm}{\ding{193}}. But  since $\taus\ll\tauphi$, the fluid-velocity gradients vary more rapidly than the inertial torque. Therefore they can be approximated by white noise, as assumed by Klett.
Using the asymptotic forms of the resistance coefficients for nearly spherical particles, we find that the condition $\taud\ll\taus$ is difficult to satisfy, it requires $\sqrt{4480/811}\ll\sv|\asp-1|\ll 4480/(811\ell)$.

In this white-noise limit we obtain that $\langle \delta\varphi^2\rangle \sim \sv^{-3}$. This result differs from  Eq.~\eqnref{klett} by a factor of  $\sv^{-1}$. The missing factor comes from the fact that the time scale of the fluid-velocity gradients  is $\taus$ for large $\sv$, not $\tauK$. As a consequence the variance catches the additional factor $\sv^{-1}$.

In  regime \raisebox{-0.2mm}{\ding{195}}, on the other hand, the tilt-angle variance is proportional to $\sv^{-2}$  in the statistical model, just
like Eq.~(\ref{eq:klett}). But the angular dynamics is driven by slip-velocity fluctuations, the Jeffery torque \eqnref{jeffery} does not matter. This   leads to a different parameter dependence
of the prefactor. In dimensional variables, our statistical-model result for regime \raisebox{-0.2mm}{\ding{195}} [\Eqnref{phiSqrPlateletsLargeSv}] reads
$\langle \delta\varphi^2\rangle \sim {\rm Re}_\lambda {\sqrt{\mathscr{E}\nu}}/{W^2}$.

In summary there are three difficulties with Eq.~(\ref{eq:klett}). First, it accounts for particle inertia in the angular dynamics but not for translational particle inertia. Our analysis shows that translational particle inertia cannot be neglected in general,
only if also rotational inertia is negligible.
Second, Eq.~(\ref{eq:klett}) assumes that the stochastic driving is isotropic white noise.
When $\taud\ll\taus$, the  fluid-velocity gradients seen by the particle can be approximated by white noise, but their diffusion time scale is given by $\taus$, not $\tauK$.
Third, this white-noise limit is difficult to achieve  unless $\asp$ is close to unity.

\subsection{Estimates of dimensionless parameters}
\label{sec:estimates}
\begin{figure}
\begin{overpic}[width=7cm,clip]{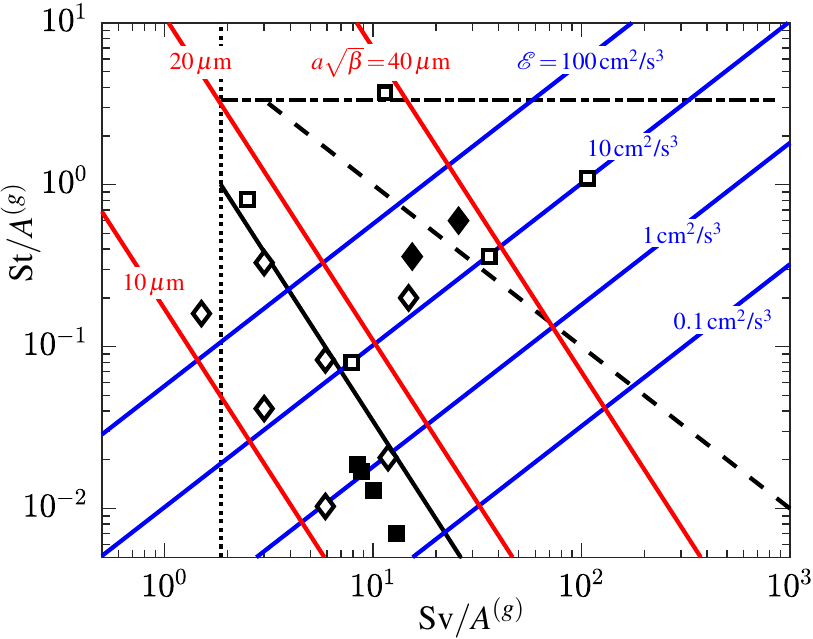}
\end{overpic}
\caption{\label{fig:params} Phase diagram, similar to Fig.~\ref{fig:PhaseDiagram}
for platelet-shaped crystals, for $\ell=10$.
Symbols show the values of the dimensionless parameters corresponding to experimental and numerical studies of non-spherical crystals  settling in turbulence (details in Supplemental Material). Dimensionless parameters  estimated from: \citep{Bre04} ($\Box$);
numerical study of collisions between disks settling in turbulence \citep{Jucha2018} ($\Diamond$);  experiments
by \cite{Kramel} ($\blacksquare$), and \cite{esteban2020disks} ($\blacklozenge$).
Blue solid lines show contours of constant turbulent dissipation rate, $\mathscr{E}=0.1,1,10,100\,$cm$^2$/s$^3$, using Eq.~(\ref{eq:froude}) with $\nu=0.1\,$cm$^2$/s, $\rho_{\rm p}/\rho_{\rm f}=1000$, and $g=980$cm/s$^2$. Red solid lines show contours of constant $a\sqrt{\beta}=10,20,40\,\mu$m  (see text).
}
\end{figure}

Parameter values for different experimental and theoretical studies of non-spherical, platelet-shaped crystals settling in turbulence are shown in Fig.~\ref{fig:params}. Note that the locations of the boundaries $\tau_\varphi=1$ and $\tau_\varphi=\taud$ depend on $\asp$, but only weakly unless $\asp$ is close to unity. The boundaries
are drawn for $\beta = 0.1$ in Fig.~\ref{fig:params}, but for slender platelets  the precise value of the aspect ratio does not matter as far as
the phase boundaries are concerned.

The boundaries of regimes \raisebox{-0.2mm}{\ding{193}} and~\raisebox{-0.2mm}{\ding{195}} are affected by the value of the turbulent Reynolds number $\ReLambda$,  because it determines the correlation length $\ell$, as explained above.
Since $\taus$ depends on $\ell$, a larger value of $\ell$ restricts
regime~\raisebox{-0.2mm}{\ding{195}} to yet larger values of $\sv$.
Moreover, since $\langle u_2^2\rangle\sim\ell^2\langle A_{21}^2\rangle$, a larger value of $\ell$ reduces the prefactor of the result in regime~\raisebox{-0.2mm}{\ding{193}}, where flow-gradients dominate, compared to the result in regime~\raisebox{-0.2mm}{\ding{194}}, where the fluid velocity dominates.
It is likely that $\ell$ is of the same order for the different data sets, but not the same. We simply set $\ell=10$ in Fig.~\ref{fig:params}.

Which values of the dimensionless parameters $\st$ and $\sv$ are relevant for ice-crystal platelets in clouds? Both $\sv$ and $\st$ are proportional
to $\tau_p$, see Table~\ref{tab:dimless}, which in turn depends on the size and the shape of the platelet through the
product $a \sqrt{\asp}$, on the turbulent energy dissipation rate $\mathscr{E}$, as well as upon the fluid viscosity $\nu$, mass-density ratio $\rho_{\rm p}/\rho_{\rm f}$, and
the gravitational acceleration $g$.
Using typical values for ice crystals in clouds, $\nu \approx 0.1\,$cm$^2$/s, $\rho_{\rm p}/\rho_{\rm f} \approx 1000$, and $g \approx 980$cm/s$^2$,
we can express $\st$ and $\sv$ in terms of the dimensional parameter
combinations $\mathscr{E}$ and $a\sqrt{\beta}$ as:
\begin{align}
\label{eq:froude}
\st\approx 8200\,(a\sqrt{\beta})^2\sqrt{\mathscr{E}}
\mbox{ and }
\sv\approx 0.0018\,\mathscr{E}^{3/4}\st\,.
\end{align}
Fig.~\ref{fig:params} shows four blue lines  corresponding to fixed values of turbulence intensity $\mathscr{E}$,  covering the range observed in
clouds~\citep{grabowski1999comments}, and three red lines at three fixed values
of $a \sqrt{\asp}$ (for platelets this is the only dependence on the value of $\beta$ because $\AgNew$ is approximately constant for $\asp < 0.1$, $\AgNew\approx 0.85$, see Table S1 in the Supplemental Material).

We took parameter values relevant to platelets settling in turbulent flows from~\cite{Bre04,Jucha2018}, empty symbols in \Figref{params}.
The corresponding Stokes number ranges from $\st/\AgNew \approx 10^{-2}$ to $3$, and the settling number ranges from $\sv/\AgNew \approx 1$ to $100$.
So the first conclusion of our analysis is that particle inertia matters
in a large range of the parameter space, not only for platelets settling in clouds, but also for other experiments: Fig.~\ref{fig:params} also shows values of $\st$ and $\sv$ from recent
experiments measuring non-spherical particles settling in turbulent water {\citep{Kramel,esteban2020disks}}.

Note  also that the lines of constant $a\sqrt{\asp}$ in Fig.~\ref{fig:params} are parallel,
in the log-log representation of the Figure, to the phase boundary between regimes
\raisebox{-0.2mm}{\ding{193}} and
\raisebox{-0.2mm}{\ding{194}}. Therefore  only the smallest platelets, with $a\sqrt{\asp} \ll 18 \mu m$,
exhibit overdamped dynamics (regime \raisebox{-0.2mm}{\ding{193}}).

In conclusion,
many of the relevant parameter values lie in the centre of the parameter plane where the different asymptotic regions meet. In these cases, the tilt-angle variance is determined by a combination of different mechanisms, and we do not expect its dependence upon the settling velocity or particle size to be of power-law form. This is the second main conclusion of our analysis.

\subsection{Comparison with observations}
\label{sec:comparison}
Our analysis shows how the tilt angle depends on particle size and on the
turbulent dissipation rate $\mathscr{E}$.
The results shown in Fig.~\ref{fig:PhaseDiagram} indicate that the
tilt angle depends quite weakly upon particle shape $\asp$ at constant $\st/\AgNew$ and $\sv/\AgNew$.

Observations~\citep{Noel:04,Noel:05} indicate that crystals
settling in cirrus clouds align very well, with fluctuations
$\langle \delta \varphi^2 \rangle^{1/2}$ no larger than
$\approx 1^o$, or equivalently
$\log_{10} \langle \delta \varphi^2 \rangle < -3.5$.
Our Fig.~\ref{fig:PhaseDiagram} shows that such small values of
$\langle \delta \varphi^2 \rangle$ can only  occur at very small
turbulence intensities,  of the order of $\mathscr{E}\sim 1\,$cm$^2$/s$^3$ or smaller.
Available measurement of the kinematic energy dissipation in cirrus clouds
is consistent with this conclusion: values in the range $0.1 - 1\,$cm$^2$/s$^3$
are observed in such clouds~\citep{Gultepe:1995,Westbrook:2010}.

Fig.~9 of \cite{Bre04} indicates that typical tilt angles of quite large ice-crystal platelets ({$a\sim$}$1\,$mm) at reasonably high cloud-turbulence levels, with $\mathscr{E}\sim 1000\,$cm$^2$/s$^3$, are of the order of a few degrees. Our  model (Fig.~\ref{fig:DNSvariance}) predicts that the variance ranges from $\langle \delta\varphi^2\rangle\sim 10^{-3}$ rad$^2$ for small Stokes numbers to $\sim 0.1$ rad$^2$ for larger Stokes numbers, corresponding to typical tilt angles between $2$ and $18$ degrees. So at larger Stokes numbers our model gives a much higher tilt-angle  than the average  estimated by \cite{Bre04}.
Thus, contrary to the predictions of~\cite{Cho81}, our results point to
a strong effect of turbulence upon alignment.
How can our estimate for the tilt-angle variance be reconciled with the much smaller one of \cite{Bre04}?
One possibility is that crystals strongly align only in
regions where the turbulence intensity is much weaker. This might explain the relatively low fraction of crystals observed
to align. Another possibility is that our model becomes inaccurate for the relatively large crystals considered by \cite{Bre04}.
We discuss the limitations of the model next. 

\subsection{Limitations of the model}
\label{sec:limitations}
The model equations assume that $\rep$ remains small, because we neglected $\rep$-corrections to the c.o.m.~dynamics
and considered only the lowest-order $\rep$-expression for the convective inertial torque, assuming that $\rep$ is less than unity.
The parameter values of \cite{Bre04} corresponding to the largest Stokes numbers have particle Reynolds numbers larger than $10$,
and the experiments of \cite{esteban2020disks} correspond to still larger particle Reynolds numbers (all parameter values are summarised in the Supplemental Material).
In our numerical computations using DNS of turbulence we kept the linear $\rep$-corrections to the c.o.m. dynamics.
The results  indicate that these corrections do not make a qualitative difference in regimes  \raisebox{-0.2mm}{\ding{193}}, \raisebox{-0.2mm}{\ding{194}}, and  \raisebox{-0.2mm}{\ding{195}}, for the chosen parameters.
At large Stokes numbers (in regime \raisebox{-0.2mm}{\ding{196}}), by contrast, our simulations show that the
correction  (\ref{eq:drag_correction}) can make a substantial difference. At present we do not know how to describe this effect in regime \raisebox{-0.2mm}{\ding{196}}.

Higher-order $\rep$-corrections to the convective  inertial torque are known in closed form only for slender columns \citep{Kha89,Lop17}.
\cite{jiang2021} quantified how well Eq.~(\ref{eq:torque_fluid_inertia}) works at larger $\rep$, for spheroids in a steady homogeneous flow. For particle Reynolds
numbers up to $\rep$ of the order of $10$  the angular dependence remains accurate, but the numerical prefactor is smaller than
predicted by Eq.~(\ref{eq:torque_fluid_inertia}). In regime  \raisebox{-0.2mm}{\ding{193}}
this leads to a larger variance.
In regime  \raisebox{-0.2mm}{\ding{196}}, by contrast, Eq.~(\ref{eq:5}) implies that the tilt-angle variance decreases, at least when the statistical model applies. In regimes \raisebox{-0.2mm}{\ding{194}} and  \raisebox{-0.2mm}{\ding{195}} it is less clear what happens,
because the asymptotic expressions \eqnref{phiSqrPlateletsPlateau} and \eqnref{phiSqrPlateletsLargeSv} are independent of the inertial-torque amplitude $\mathscr{A}$. Finally, smaller values of the torque imply that the condition $\tau_\phi \sim 1$, which defined the transition from regime~\raisebox{-0.2mm}{\ding{192}}, is shifted to higher values of $\sv$, extending the parameter range where the tilt angle is uniformly distributed.

As $\rep$ increases, the dynamics of disks settling in a quiescent fluid  becomes unstable \citep{auguste2013falling,esteban2020disks}, because the symmetry of the disturbance flow is broken, and because it becomes unsteady. Our model cannot describe these effects due to fluid nertia.

The model  uses a steady approximation for the convective inertial torque \citep{Kramel,Men17,Lop17,Gus19,Sha19}. The experiments
by \cite{Lop17} indicate that this is at least qualitatively correct for rods settling in a cellular flow, although the slip velocity $\ve W(t)$ fluctuates as a function of time. In general, however, the steady model must break down when $\ve W(t)$ fluctuates too rapidly
\citep{candelier2019time}. For the steady model to hold  in our case, it is necessary that $\taus$ is much larger than the viscous time,  $\taus \gg a^2/\nu$ (in this Section we use dimensional variables). Otherwise unsteady effects may arise, analogous to the Basset-Boussinesq-Oseen history force for the c.o.m.~motion of a sphere in a quiescent fluid. For the cellular flow with correlation length $\ell$ \citep{Lop17}
we require $\ell/W \gg a^2/\nu$. Using $\rep = Wa/\nu$ this means $\ell/(\rep a) \gg 1$.
In the experiments, $\ell  \sim 1\,$cm,  $a\sim 1\,$mm, and the largest Reynolds number
is $\rep \sim 10$. So the condition is marginally satisfied for the largest  $\rep$.
In the statistical model, using $\taus/\tauK = (\AgNew/\sv) (\ell/\etaK)$  (in dimensional variables), the condition translates to $\sv \ll  \AgNew\ell \etaK/a^2$, consistent with the constraint (\ref{eq:assumption2}).

The model
also neglects the convective-inertial torque due to fluid shears \citep{subramanian2005,einarsson2015a,rosen15}.
This is justified  if $a/\etaK\ll 1$ \citep{Candelier2016}, but for larger particles the shear-induced torque might matter. This torque has a different physical origin from the convective-inertial torque due to finite slip. The former is determined by the disturbance flow close to the particle,
while the latter is due to far-field effects, where the presence of the particle is approximately taken into account by a singular source term.
As a first approximation, one could therefore simply superimpose the torques due to shear and due to slip.

Our model assumes that the ice crystals are homogeneous, in other words that the mass
density is the same throughout the crystal.
With this constraint, the choice of model parameters  is overall
consistent with known properties of ice crystals in clouds.
The values of $\asp$ chosen in Fig.~\ref{fig:params} for crystals
of diameter $300 \mu m$ (empty diamond symbols), $0.01 \le \beta \le 0.05$,
lead to
settling velocities consistent with those reported in laboratory
studies~\citep{Kajikawa:72,Pru78}. The empty square symbols are
taken from the study of~\cite{Bre04}, and also correspond to realistic
settling velocity of crystals, consistent with known
results~\citep{Auer:70,Heymsfield1972}.

In reality,
the microscopic growth processes of cloud crystals may result in inhomogeneous mass densities \citep{Heymsfield1973}. The crystals may even
contain hollow regions and may exhibit irregular shapes \citep{Korolev2000,Heymsfield2002b}.
How such imperfections affect the dynamics of ice crystals is not considered in our model.
An inhomogeneous mass distribution results in an additional gravitational torque which could affect the angular dynamics.
This is well studied for nearly neutrally buoyant marine microorganisms settling in the turbulent ocean \citep{Kes85,Cen13,Gus15b}, but little is known for heavy particles
settling in air.
Shape irregularities  can affect the inertial contribution to the  torque \citep{Kha89,Can16}.
The highly idealised model of \cite{Can16} shows that such asymmetries have a negligible effect on the dynamics of slender columns if
 $\delta a_\perp /a_\perp \ll \beta{\rm Re}_{\rm p}$, their Eq. (4.11). The experiments and the refined analysis
 of \cite{roy2019symmetry} bear out this qualitative prediction. Conversely, if $\delta a_\perp/a_\perp\gg\beta{\rm Re}_{\rm p}$, then the asymmetry dominates the angular dynamics. Slender columns settle vertically in this limit, aligned with gravity.
 This is  not observed for columnar crystals in clouds.  

\section{Conclusions}
\label{sec:conclusions}
Particle inertia increases the tilt-angle variance of small crystals settling through a turbulent cloud because it gives rise to additional fluctuations in the angular equation of motion.
Even at  very small Stokes numbers this can be a significant effect, since  the overdamped theory \citep{Kramel,Men17,Gus19,Anand2020} predicts a very small variance. For neutrally buoyant particles the overdamped theory works fairly well. But for ice crystals
that are about 1000 times heavier than air  it
 can underestimate the variance by a large factor.
 Moreover, we found that particle inertia matters in a large region in parameter space (Fig.~\ref{fig:params}).

The problem has four relevant time scales (Table \ref{tab:TimeScales}). As a consequence there are many different asymptotic regimes where the tilt-angle variance displays different dependencies on the dimensionless parameters (Table \ref{tab:dimless}), in particular different power laws
as a function of the settling number $\sv$. Relevant dimensionless parameters tend to lie in a central region in the parameter plane where the different asymptotic regimes meet, so that the tilt-angle variance is determined by a combination of different physical mechanisms. In this case there are no simple power-law dependencies on the settling velocity [Eqs.~(\ref{eq:klett}) or (\ref{eq:od1})].

Our results are  based on a small-angle expansion, as first used by \cite{Kle95} for this problem. Other assumptions of his theory are not satisfied in the regimes we studied, so that its main prediction (\ref{eq:klett}) does not describe our simulation results.

Our analysis shows that the very strong alignment of ice crystals
in cirrus clouds, with typical tilt angles $\langle \varphi^2 \rangle^{1/2}\sim 1^o$ or less~\citep{Noel:04,Noel:05}
is only possible when
the turbulence intensity if low, of the order of $0.1-1\,$cm$^2$/s$^3$. This
is generally consistent with the known value of turbulent energy
dissipation rate  in such ice clouds~\citep{Gultepe:1995}. In fact, an important
future test of our theory is to correlate the presence of strongly
oriented ice crystal~\citep{Noel:10} with the local level of cloud turbulence.
Another test is to calculate what a satellite sensor would detect,
starting from the size distribution of either columns or platelets obtained by aircraft observations~\citep{Heymsfield2002b},
together with estimates of local cloud-turbulence levels.

Such tests are important because the model was derived under a number of assumptions. First, we assumed that the particle Reynolds number  is small.
Second, we assumed that the torque is obtained by simply superimposing the fluid-inertia torque and the Jeffery torque. But as we discussed,
there are additional contributions to the torque when turbulent shears give rise to convective fluid inertia.
For crystals smaller than the Kolmogorov length, these contributions are negligible because
the shear Reynolds number is small \citep{Candelier2016}, but for larger crystals they may become important.
Third, Eq.~(\ref{eq:torque_fluid_inertia}) was derived in the steady limit.
For the steady model to hold it is necessary that the fluctuations of the slip velocity are slow compared
with the viscous time. At very large settling numbers this constraint is broken. It remains a question for the future to describe the effect of unsteady torques.

In our discussion of the results we focused on the variance of the tilt angle $\varphi$, the angle between the particle-symmetry vector and the direction of gravity. But to compute the effect of particle inertia we needed to consider a second angle, $\theta$, that describes how the particle-symmetry vector rotates in the plane perpendicular to the direction of gravity. Regarding the dynamics of $\theta$ we found significant differences between columns and platelets, described in the Supplemental Material. For the tilt-angle variance these differences do not matter much, but they are likely to be important for collisions between ice crystals, a question that  remains to be explored \citep{aurore}.

We remark that although this study focused on the variance of the tilt angle, the method leading to \Eqnref{phiSqrPlateletsUnevaluated} and outlined in the Supplemental Material, can straightforwardly be extended to calculate higher-order moments or correlation functions of the tilt angle.

When particle inertia becomes important, preferential sampling may affect the statistics of observables such as the tilt angle. This is well known for heavy spherical particles in turbulence \citep{Gus16}.
Our results show that preferential sampling is a weak effect, at least for the parameters considered here.

Finally, we assumed that the particles are much heavier than the fluid, this is the limit relevant for ice crystals in clouds. But recent experimental studies
 \citep{Kramel,Lop17} used nearly neutrally buoyant particles. In this case one expects the effect of particle inertia to be  weaker. It remains an open question under which circumstances particle inertia may nevertheless make a noticeable difference.

%

\acknowledgments
BM thanks Fabien Candelier for discussions regarding the inertial torque. KG and BM were supported by the grant {\em Bottlenecks for particle growth in turbulent aerosols} from the Knut and Alice Wallenberg Foundation, Dnr. KAW 2014.0048, and in part by VR grant no. 2017-3865 and Formas grant no. 2014-585.  AP and AN acknowledge support from the IDEXLYON project (Contract ANR-16-IDEX-0005) under University of Lyon auspices.  Computational resources were provided by C3SE and SNIC, and PSMN.


\end{document}


\title{Supplemental Material for \lq Effect of particle inertia on the alignment of small ice crystals in turbulent clouds\rq{}}
\author{K. Gustavsson$^1$, M. Z. Sheikh$^2$, A. Naso$^3$, A. Pumir$^2$, B. Mehlig$^1$}
\affiliation{\mbox{}$^1$ Department of Physics, Gothenburg University, 41296 Gothenburg, Sweden\\
\mbox{}$^2$Univ. Lyon, ENS de Lyon, Univ. Claude Bernard, CNRS, Laboratoire de Physique, F-69342, Lyon, France\\
\mbox{}$^3$Univ. Lyon, Ecole Centrale de Lyon, Univ. Claude Bernard, CNRS, INSA de Lyon, Laboratoire de M\'ecanique des Fluides et d'Acoustique, F-69134, Ecully, France}

\maketitle

\section{Resistance and inertia tensors for a spheroid}
\label{sec:resistance}
For a spheroid, the resistance tensors in Eqs.~(9) and (10) in the main text, and the particle
inertia tensor in Eq.~(8) are given by \cite{Kim:2005}. For convenience we summarise the relevant formulae in this Section.
The translational resistance tensor $\ma A$ of a spheroid reads
\begin{equation}
A_{ij} \equiv
A_\perp (\delta_{ij}-n_i n_j) + A_\parallel n_i n_j\,,
\label{eq:def_Mt}
\end{equation}
with  coefficients
\begin{align}
A_\perp&=\frac{8(\asp^2-1)}{3\asp[(2\asp^2-3)\gamma+1]}\,,\quad
A_\parallel=\frac{4(\asp^2-1)}{3\asp[(2\asp^2-1)\gamma-1]}\,,\quad
\gamma=\frac{\ln[\asp + \sqrt{\asp^2-1}]}{\asp\sqrt{\asp^2-1}}\,.
\end{align}
These expressions are consistent with those given in Tables~3.4 and 3.6 \new{in the book} by \cite{Kim:2005}. For oblate spheroids, our expressions
are identical to those of Kim \& Karrila, for prolate spheroids they differ by a factor of $\asp$, and so does Eq.~(9) in the main text. As a consequence, the two formulations are equivalent for oblate and for prolate spheroids.
In the main text we refer to the translational resistance \new{coefficients} in the $\hat{\ve g}$- and $\hat{\ve p}$-directions (\new{these directions are defined in }Fig.~2 in the main text). \new{These} coefficients are defined as
\begin{align}
\AgNew\equiv\left\{
\begin{array}{ll}
\!\Ao & \mbox{\!\!\!if }\Lambda>0\cr
\!\Ap & \mbox{\!\!\!if }\Lambda<0
\end{array}
\right.
\!,\;\;
\ApNew\equiv\left\{
\begin{array}{ll}
\!\Ap & \mbox{\!\!\!if }\Lambda>0\cr
\!\Ao & \mbox{\!\!\!if }\Lambda<0
\end{array}
\right.
\,.
\eqnlab{AxNewDef}
\end{align}
\new{Recall} that $\Lambda > 0 $ ($\Lambda < 0$) corresponds to prolate
(oblate) spheroids.
 \new{The}  resistance tensors $\ma C$ and $\ma H$ in the Jeffery torque [Eq.~(10) in the main text] take the form:
\begin{align}
C_{ij} &\equiv
C_\perp (\delta_{ij}-n_i n_j) + C_\parallel n_i n_j\,,
\quad H_{ijk} =H_0\epsilon_{ijl} n_k n_l\,,
\label{eq:def_Mr}\\
C_\perp&=\frac{8a_\parallel a_\perp(\asp^4-1)}{9\asp^2 [(2\asp^2-1)\gamma-1]}\,,\quad C_\parallel=-\frac{8a_\parallel a_\perp(\asp^2-1)}{9(\gamma-1)\asp^2}\,,\quad
H_0= -C_{\perp}\frac{\asp^2 - 1}{\asp^2 + 1}\,.
\nonumber
\end{align}
Here $\epsilon_{ijl}$ is the Levi-Civita tensor, and repeated indices are summed  over.
The particle-inertia tensor per unit mass, $\ma I$, has elements
 \begin{align}
 \label{eq:inertiatensor}
& \hspace*{-2mm} I_{ij} =I_\perp (\delta_{ij}-n_i n_j) + I_{\parallel} n_i n_j
\,,\quad\mbox{with}\quad
I_{\perp}=\frac{1+\asp^2}{5}a_\perp^2
\quad\mbox{and}\quad
I_{\parallel}=\frac{2}{5}a_\perp^2\,.
\end{align}

\section{Calculation of the tilt-angle variance in regimes \protect\raisebox{-0.2mm}{\protect\ding{194}}, \protect\raisebox{-0.2mm}{\protect\ding{195}}, and \protect \raisebox{-0.2mm}{\protect\ding{196}}}
In this Section we describe how we determined the tilt-angle variance  $\langle \dphi^2\rangle$ in regimes  \raisebox{-0.2mm}{\ding{194}},  \raisebox{-0.2mm}{\ding{195}}, and  \raisebox{-0.2mm}{\ding{196}}. Here $\sv$ is large, and in regimes \raisebox{-0.2mm}{\ding{194}} and \raisebox{-0.2mm}{\ding{195}}
the Stokes number is small. In regime \raisebox{-0.2mm}{\ding{196}}, the Stokes number need not be small, but we assume that $\delta\varphi$ remains small.
In regime \raisebox{-0.2mm}{\ding{194}}, the angular dynamics is overdamped. We discuss the overdamped angular dynamics in Sections \ref{sec:odp}  and  \ref{sec:odc}.
 Section \ref{sec:4} describes the angular dynamics in regime \raisebox{-0.2mm}{\ding{195}}.
In Section \ref{sec:translational} we show how to approximate over- and underdamped c.o.m.~dynamics, and \new{in Section \ref{sec:tav}} we summarise how to compute the tilt-angle variance in regimes \raisebox{-0.2mm}{\ding{193}} to \raisebox{-0.2mm}{\ding{195}}. In Section \ref{sec:5} we discuss the tilt-angle variance in regime \raisebox{-0.2mm}{\ding{196}}, where both c.o.m. and angular dynamics are underdamped.

\subsection{Overdamped angular dynamics: platelets}
\label{sec:odp}
Let us assume that the matrix $\ma Y$ \new{[Eq.~(19)} in the main text\new{]} changes slowly compared to the \new{linearized dynamics [Eq.~(18a) in the main text] and that it therefore can be treated as constant at those time scales}. For constant $\ma Y$ we find that the angular dynamics\new{, Eq.~(18a), has} the fixed point
\begin{align}
\eqnlab{FPplatelet}
&[\dphi^*,\theta^*,\omega^*_s,\omega_p^*,\omega_g^*]
=\bigg[\frac{\sqrt{Y_{g2}^2+Y_{g3}^2}}{\mbox{sgn}(\dphi)Y_{gg}},{\rm atan}\left(\frac{Y_{g3}}{Y_{g2}}\right),0,0,\Omega_g\bigg]\,.
\end{align}
The five eigenvalues of the stability matrix of the dynamics (18a) \new{evaluated} at the fixed point read
\begin{align}
\lambda_{1,2}=-\frac{\Co}{2\Io\st}\pm{\rm i}\Omega_0\,,\;\lambda_{3,4}=\lambda_{1,2}\,,\;\lambda_5=-\frac{\Cp}{\Ip\st}\,,
\new{\mbox{ where }\Omega_0=\frac{\Co}{2\Io\st}\sqrt{4|\mathscr A|\sv^2\Io\st\frac{\ApNew}{\Co\AgNew}-1}\,.}
\eqnlab{eigenvaluesPlatelet}
\end{align}
This shows that the fixed point is stable,
so that the angular dynamics relaxes to this fixed point on the time scale $2\Io\st/\Co\sim \taud$ ($\Ip\st/\Cp$ is of the \new{order $\taud$} for spheroids, unless
$\asp$ is very large). In the limit of large $\sv$, the matrix $\ma Y$ changes on the time scale $\taus$, Eq.~(20) in the main text.
We therefore expect the dynamics to follow \Eqnref{FPplatelet} when $\taud\ll \taus$. In this limit the angular dynamics is overdamped,
so that only the terms inversely proportional to $\st$ matter in Eq.~(18a).
In addition we need to assume that $\tauphi\ll 1$ to ensure that $\delta\varphi$ remains small, so that
the linearisation used to derive Eq.~(18a) remains valid.
\new{The assumptions above define the boundaries of regimes \raisebox{-0.2mm}{\ding{193}} and \raisebox{-0.2mm}{\ding{194}} given in Section~5 of the main text. }

We add a remark concerning the dynamics of $\hat{\ve p}$. To derive the expression for $\theta^*$ in \Eqnref{FPplatelet}, we solved $Y_{gs}=\gghat\T\ma Y\hat{\ve s}=0$ \new{[from the equation for $\dot\omega_p$ in Eq. (18a)]}.
For this condition to hold, $\theta^*$ must be chosen such that the vector $\gghat\T\ma Y$ is perpendicular to $\hat{\ve s}$.
Since $\hat{\ve p}$ is perpendicular to $\hat{\ve s}$, we find $\theta^*$ through the relation
\begin{align}
\hat{\ve p}(\theta^*)={\frac{{\rm sgn}(\delta\varphi)}{\sqrt{{Y_{g2}}^2+{Y_{g3}}^2}}\left[0,Y_{g2},Y_{g3}\right]}\,,
\eqnlab{overdamped_pstar}
\end{align}
for $\gghat=\hat{\ve e}_1$.
The sign is chosen so that the \new{factor} $\mbox{sgn}(\dphi^*)$ \new{in \Eqnref{FPplatelet}} equals the sign of $\dphi$ (\new{note that }$Y_{gg}$ is positive for large settling velocities). This choice  corresponds to the stable fixed point of the $\theta$-dynamics.
We conclude that the transversal orientation $\hat{\ve p}$ of platelets aligns with the vector
$[0,Y_{g2},Y_{g3}]$.

\subsection{Overdamped angular dynamics: columns}
\label{sec:odc}
Now consider the fixed points of the angular dynamics (18b) for columns.
\new{They exist only} for certain $\ma Y$-matrices, namely when there is a  $\theta^*$ such that $s_i[Y_{gi}Y_{gj}/Y_{gg}-Y_{ij}]p_j|_{\theta=\theta^*}=0$.
For large $\sv$, we approximate $Y_{gi}Y_{gj}/Y_{gg}-Y_{ij}\sim O_{ij}+\Lambda S_{ij}$. This means that fixed points of Eq.~(18b)  must satisfy $\ve s\T[\ma O+\Lambda \ma S]\ve p|_{\theta=\theta^*}=0$.
This equation \new{has} real solutions $\theta^*$ \new{if} the discriminant $\Lambda^2(S_{22}-S_{33})^2 + 4\Lambda^2 S_{23}^2 - 4 O_{23}^2$ of the lower right $2\times 2$ block of $\ma O+\Lambda \ma S$ is positive. Moreover, even when such fixed points exist, they have a vanishing stability exponent in the $\theta$-direction within the approximation considered here. Fixed points of the dynamics (18b) are therefore at best weakly \new{stable and} the dynamics does not have time to approach any fixed point within the time scale \new{of $\ma Y$, as was the case for platelets.}

Instead, we search for fixed points of the dynamics of $[\dphi,\omega_s,\omega_p,\omega_g]$.
Solving $\tfrac{\rm d}{{\rm d}t}\dphi=\tfrac{\rm d}{{\rm d}t}\omega_s=\tfrac{\rm d}{{\rm d}t}\omega_p=\tfrac{\rm d}{{\rm d}t}\omega_g=0$ for general values of $\theta$ we find:
\begin{align}
[\dphi^*,\omega^*_s,\omega_p^*,\omega_g^*]=\bigg[\frac{Y_{gp}}{Y_{gg}},0,\Omega_p,\frac{Y_{gp}Y_{gs}}{Y_{gg}}- Y_{sp}\bigg]\,.
\eqnlab{FPcolumn}
\end{align}
\new{This fixed point is stable because its} stability exponents have negative real parts,
\begin{align}
\lambda_{1,2}=-\frac{\Co}{2\Io\st}\pm{\rm i}\Omega_0\,,\;\lambda_{3}=-\frac{\Co}{\Io\st}\,,\;\lambda_4=-\frac{\Cp}{\Ip\st}\,.
\end{align}
\new{What} about the $\theta$-dynamics? Since $\tfrac{\rm d}{{\rm d}t}\theta=\omega_g\sim\omega_g^*\sim \Omega_g+\Lambda S_{sp}$ for large $\sv$, the variable $\theta$ diffuses slowly in this limit.
The stability time of the $\theta$-dynamics is  $\sim\tauK = 1$, so that $\theta$ cannot follow the fixed point.
In homogeneous, isotropic flows,  the gradients $\Omega_g+\Lambda S_{sp}$ are independent from $O_{gp}+\Lambda S_{gp}$ and $O_{gs}+\Lambda S_{gs}$, evaluated
at the same spatial position (these two combinations drive the other components of the angular velocity). Therefore we can treat $\theta$ as an independent, uniformly distributed random variable that can be considered constant  in regimes   \raisebox{-0.2mm}{\ding{193}}
and  \raisebox{-0.2mm}{\ding{194}}.
In summary, the $\delta\varphi$-dynamics of columns follows the fixed point \eqnref{FPcolumn},
and $\theta$ (and thus $\hat{\ve p}$) changes randomly albeit very slowly.

\subsection{Angular dynamics in regime \raisebox{-0.2mm}{\ding{195}}}
\label{sec:4}
In regimes \raisebox{-0.2mm}{\ding{193}} and \raisebox{-0.2mm}{\ding{194}}, the flow decorrelation time due to settling, $\taus$, is larger than the time scale of the angular dynamics. Therefore the particle orientations follow the fixed points \eqnref{FPplatelet} and \eqnref{FPcolumn}. In regime \raisebox{-0.2mm}{\ding{195}}, by contrast, $\taus$ is smaller than $\taud$.
Surprisingly, we find that the variables determining the tilt-angle dynamics -- $[\dphi,\,\theta,\,\omega_s,\,\omega_p]$ for platelets and $[\dphi,\,\omega_s]$ for columns -- nevertheless follow the fixed points, Eqs.~\eqnref{FPplatelet} and \eqnref{FPcolumn}.
This is explained by the observation that the dynamics of $[\omega_s,\,\omega_p]$ (platelets) and $\omega_s$ (columns) is fast, it fluctuates on a time scale  much shorter than $\taus$. In a mean-field approximation, we can therefore replace $\omega_s$ and $\omega_p$ in Eq.~(18) in the main text by their \new{mean values, $\langle \omega_p\rangle=\langle\omega_s\rangle=0$}. The \new{resulting mean-field equations obtained by averaging $\dot\omega_s$ and $\dot\omega_p$} become
\begin{subequations}
\label{eq:mf}
\begin{align}
\label{eq:mf1}
&Y_{gp} - Y_{gg}\dphi=0\quad\mbox{and}\quad Y_{gs}=0\quad\mbox{for platelets}\,,\\
&Y_{gp} - Y_{gg}\dphi=0\quad\mbox{for columns}\,.
\end{align}
\end{subequations}
In Eq.~(\ref{eq:mf1}) we neglected the terms quadratic in $\omega$ in Eq.~(18a) in the main text. Since the sign of $\delta\varphi$ does not fluctuate, these terms have non-zero \new{averages}.
However, computer simulations of the statistical model show that these terms remain bounded, even though they are proportional to $\delta\varphi^{-1}$. When the Stokes number is small\new{,} we therefore neglected these terms, \new{but kept terms multiplied by $\st^{-1}$. This results in} Eqs.~(\ref{eq:mf}).
These equations are the same as the conditions obtained when solving Eq.~\eqnref{FPplatelet}
 for the fixed-point
values of $\dphi^*$ and $\theta^*$ (platelets),  and \eqnref{FPcolumn} for $\dphi^\ast$ (columns).
We therefore conclude that the solutions for  $\dphi^*$ (and $\theta^*$ for platelets) obtained in regimes
\raisebox{-0.2mm}{\ding{193}} and \raisebox{-0.2mm}{\ding{194}} remain valid also in regime \raisebox{-0.2mm}{\ding{195}}. But note that other components of $\ve\omega$ cease to follow their fixed-point values in regime \new{\raisebox{-0.2mm}{\ding{195}}}.

For all simulations shown in the \new{figures} in the main text we verified that the dynamics  of $[\omega_s,\,\omega_p]$ (platelets) and $\omega_s$ (columns) is fast. For columns we can show this explicitly in regime \raisebox{-0.2mm}{\ding{195}} where we can replace $W_p$ \new{in Eq.~(18)} by $-u_2$ \new{because fluctuations of $\ve v$ are much smaller than fluctuations of $\ve u$ in the underdamped limit}.  For columns we infer that $\dphi$ satisfies the equation of a driven damped oscillator
\begin{align*}
\tfrac{{\rm d}^2}{{\rm d}t^2}\dphi+\tfrac{\Co}{\Io\st}\tfrac{{\rm d}}{{\rm d}t}\dphi+\tfrac{\Co}{\Io\st}|{\mathscr A}|\sv^2\frac{\ApNew}{\AgNew}\dphi= -|{\mathscr A}|\sv\ApNew\tfrac{\Co}{\Io\st}u_{2}\,.
\end{align*}
The solution of this equation is
\begin{align}
\label{eq:dho2}
\dphi(t) = \frac{\AgNew\Omega_0}{\sv}\int_0^t{\rm d}t_1e^{(t_1-t)\Co/(2\Io\st)} \sin[\Omega_0(t_1-t)]u_2(t_1)\,,
\end{align}
where \new{$\Omega_0=\sqrt{|\mathscr A|\sv^2\Co\ApNew/(\Io\st\AgNew)}$ [$\Omega_0$ in \Eqnref{eigenvaluesPlatelet} approaches this expression in the limit of} regime \raisebox{-0.2mm}{\ding{195}}].
Taking the time derivative gives $\omega_s(t)$. To show that the dynamics of $\omega_s(t)$ is faster than that of $\dphi(t)$, we compute the correlation functions $\langle\dphi(t)\dphi(t')\rangle$ and $\langle\omega_s(t)\omega_s(t')\rangle$ using \Eqnref{dho2}.
We find that $\langle\dphi(t)\dphi(t')\rangle$ decays like
$C_u(t-t')\new{=\langle u_2(t-t')u_2(0)\rangle}$ \new{and} conclude that $\dphi$ decorrelates on the time scale $\taus$.
To estimate the correlation time of $\omega_s$, we expand $C_{\omega_s}(t-t')=\langle\omega_s(t)\omega_s(t')\rangle$  for small $|t-t'|$, $C_{\omega_s}(t-t')\sim C_{\omega_s}(0)+C''_{\omega_s}(0)t^2/2$, and  define the correlation time $\tau_{\omega_s}$ of $\omega_s$ in terms of the curvature of the correlation function
\begin{align}
\label{eq:tau_omega_s}
\tau_{\omega_s}={\sqrt{-C_{\omega_s}(0)}}/{{C''_{\omega_s}(0)}}\,.
\end{align}
Using that  $\Omega_0\sim\sqrt{\tauphi\taud}$ is much larger than $\taus^{-1}$ in regime \raisebox{-0.2mm}{\ding{195}}, \Eqnref{tau_omega_s}
yields $\tau_{\omega_s}\sim \sqrt{2}/(\Omega_0^2\taus)$.
We therefore conclude for columns that the correlation time of $\omega_s$ is much smaller than the correlation time of $\dphi$  in regime \raisebox{-0.2mm}{\ding{195}}.

In summary,  we observe that the components
of $\ve\omega$ fluctuate around zero with different time scales. For platelets, $\omega_s$ and $\omega_p$ fluctuate rapidly,  the dynamics of the other components  is much slower.
Averaging over the fast dynamics we find that
$\dphi$ and $\theta$ follow the fixed point (\ref{eq:FPplatelet}).
The conclusions for platelets are slightly different.  In this case $\omega_s$ varies rapidly. Averaging over the fast dynamics, $\dphi$ approximately follows \eqnref{FPcolumn}. We also observe that $\theta$ varies slowly compared to the other variables.
Taken together  we conclude that $\delta\varphi$ continues to follow its fixed point $\dphi^\ast$ -- \Eqnref{FPplatelet} (platelets) and \Eqnref{FPcolumn} (columns) -- even in regime \raisebox{-0.2mm}{\ding{195}},
although the angular dynamics is underdamped. This means that we can use $\dphi\approx \dphi^\ast$ in regimes  \raisebox{-0.2mm}{\ding{193}},  \raisebox{-0.2mm}{\ding{194}}, and  \raisebox{-0.2mm}{\ding{195}}.
When $\taud\gg 1$,  by contrast, the dynamics is too slow to relax to these fixed points.
We derive a theory for this regime in Section~\ref{sec:5}.

\subsection{Centre-of-mass dynamics for small tilt angle $\delta\varphi$}
\seclab{translational}
Since the translational degrees of freedom enter the angular dynamics (18) through $\ma Y$ [Eq.~(19) in the main text],
we must determine how the slip velocity $\ve W$  fluctuates.
For large settling velocity  we have
\begin{align}
v_g\approx W_g\approx {\sv}/{\AgNew}\,,\quad Y_{gg}\approx |{\mathscr A}|\sv^2{\ApNew}/{\AgNew}\,.
\eqnlab{WgApprox}
\end{align}
When particle inertia matters ($\taud\gg\tauphi$), $\ve W$ no longer follows \new{$\ve W^{(0)}$}. Fluctuations of $\ve W$ around $\ve W^{(0)}$ affect the angular dynamics, and this means that Eq.~(16) in the main text, and the corresponding result for platelets fails.

To describe the fluctuations of $\ve W$ we use that the settling speed $W_g$ is large.
In this limit we can neglect preferential sampling due to deviations from the deterministic settling path \citep{Gus17}
$\xxd{t}(\hat{\ve n}_0)\equiv\ve x_0+\sv\ve W^{(0)}{(\hat{\ve n}_0)}\,t$ when evaluating the fluid velocity at the particle position, and when evaluating its spatial gradients.
Here $\ve x_0$ is the initial particle position.
Since $\dphi$ is small, we choose for the direction of $\hat{\ve n}_0$ the stable
orientation $\phizerostar$.
The fluid velocity and its spatial gradients are therefore evaluated along the path
\begin{equation}
\xxd{t}\equiv\left.\xxd{t}(\hat{\ve n}_0)\right|_{\varphi_0=\phizerostar}=\ve x_0+\frac{\sv}{\AgNew}\,\gghat\, t\,.
\eqnlab{xxdDef}
\end{equation}
If we approximate $\ve x(t) \approx \xxd{t}$, then the centre-of-mass equation of motion [Eq.~(7) in the main text] becomes to first order in $\dphi$
\begin{equation}
\label{eq:small_angle_t}
\dot v_k \!=\! \frac{\Ao\! -\! \ApNew}{\st}W_pp_k \!-\!\frac{\Ao W_k}{\st}\!+\! \frac{\dphi\sv}{\st}\Big(\frac{\ApNew}{\AgNew}\!-\!1\Big) p_k\,,
\end{equation}
for  $k=2,3$ (transversal to $\gghat$), and $\ve W=\ve v-\ve u(\xxd{t},t)$.

\subsection{Tilt-angle variance in regimes \raisebox{-0.2mm}{\ding{193}}, \raisebox{-0.2mm}{\ding{194}}, and \raisebox{-0.2mm}{\ding{195}}}
\label{sec:tav}
We start with platelets.
To obtain the corresponding dynamics of $Y_{gk}$, we differentiate Eq.~(19) in the main text w.r.t. time \new{using \Eqnref{small_angle_t}} and set $\dphi\approx\dphi^*$ and $\theta\approx\theta^*$ [Eq.~(\ref{eq:FPplatelet})]. This yields:
\begin{align}
\dot Y_{gk} =& -\frac{\AgNew}{\st}Y_{gk} - |\mathscr A|\ApNew\sv \dot u_k
 -\dot O_{1k} -|\Lambda|\dot S_{1k} - \frac{\ApNew}{\st}(O_{1k}+|\Lambda| S_{1k})
\eqnlab{YgkEqm}
\end{align}
for $k=2,3$.
Integrating this differential equation we find:
\begin{align}
\eqnlab{YgkSol}
Y_{gk}(t)&= - \ApNew|\mathscr A|\sv u_k(t) -O_{1k}(t)-|\Lambda|S_{1k}(t)
\\&
+ \frac{\AgNew}{\st}\int_0^t{\rm d}t_1 e^{(t_1-t)\AgNew/\st}\Big[\ApNew|\mathscr A|\sv u_k(t_1)
+\Big(1-\frac{\ApNew}{\AgNew}\Big)(O_{1k}(t_1)+|\Lambda|S_{1k}(t_1))\Big]\,.
\nn
\end{align}
Here and in the following we use the notation $\ve u(t) \equiv \ve u(\xxd{t},t)$, also for the fluid-velocity gradients.
Inserting Eqs.~\eqnref{YgkSol} and \eqnref{WgApprox} into \Eqnref{FPplatelet} we obtain an expression for $\dphi^*$.
We find the tilt-angle variance by squaring and averaging $\dphi^*$. Assuming homogeneous and isotropic fluid-velocity statistics yields:
\begin{align}
\eqnlab{phiSqrPlateletsUnevaluatedAppendix}
&\langle\dphi^2\rangle=
f_\Lambda\Big\{\frac{\AgNew{}^2}{\sv^2}C_u(0)
+\frac{\AgNew{}^2}{\ApNew{}^2|\mathscr A|^2\sv^4}C_B(0)
\\&
+\frac{\AgNew}{|\mathscr A|^2 \st \sv^4}\int_0^\infty {\rm d}t e^{-\AgNew t/St}\Big[\Big(1-\frac{\AgNew{}^2}{\ApNew{}^2}\Big)C_B(t)
+2\AgNew |\mathscr A| \sv C_X(t)
-\AgNew{}^2 |\mathscr A|^2 \sv^2 C_u(t)\Big]\Big\}\,,
\nn
\end{align}
where we set $f_\Lambda=2$ for $\Lambda < 0$ (platelets).
This is our  main result, Eq.~(23) in the main text,
 for the tilt-angle variance of platelets settling in a turbulent flow, expressed in terms of the correlation functions of fluid velocities and fluid-velocity gradients, $C_B(t)=\langle O_{12}(t)O_{12}(0)+2|\Lambda|O_{12}(t)S_{12}(0)+\Lambda^2S_{12}(t)S_{12}(0)\rangle$, $C_u(t)=\langle u_{2}(t)u_{2}(0)\rangle$ and $C_X(t)=\langle u_2(t)[O_{12}(0)+|\Lambda|S_{12}]\rangle$ (Section \ref{sec:eval}).

For columns the variance of $\dphi$  is obtained by averaging  $[\dphi^*]^2$ in \Eqnref{FPcolumn} over independent, uniformly distributed values of $\theta$, resulting in
\begin{align}
\langle [\delta\varphi^\ast]^2\rangle_\theta=\frac{Y_{g2}^2+Y_{g3}^2}{2Y_{gg}^2}\,.
\end{align}
We were surprised to find that this result for columns equals  $\tfrac{1}{2}[\dphi^*]^2$ for platelets, Eq.~\eqnref{FPplatelet}.
In general the dynamics of  $Y_{gk}$  for columns is  different from \Eqnref{YgkSol}.
However, the combination $Y_{g2}^2+Y_{g3}^2$ is the same for the two kinds of particles within our approximations,
and in both cases $Y_{gg}$ is given by Eq.~\eqnref{WgApprox}.
This means that $\langle\dphi^2\rangle$ takes the form of \Eqnref{phiSqrPlateletsUnevaluatedAppendix} divided by two.
We take this factor into account by defining $f_\Lambda=1$ for $\Lambda >0$ (columns).

\subsection{Tilt-angle variance in regime \raisebox{-0.2mm}{\ding{196}} }
\label{sec:5}
In this regime both the c.o.m. and the angular dynamics are underdamped. The settling velocity is large, and the Jeffery torque [Eq.~(10) in the main text] does not matter. Therefore  we
can simplify Eq.~(18) in the main text
by setting  $W_g=\sv/\AgNew$, $W_k=-u_k$, and $\ma A=0$.
We can assume that $\theta$ is an independent variable, allowing us to replace $u_p$ by $u_2$ and $u_s$ by $u_3$.
When $\taus$ is the smallest time scale, we may approximate the $\dphi$-dynamics  as Langevin equations on the form $\delta\ve X=\ve V\delta t+\delta\ve F$.
For columns, the relevant variables are $\ve X=[\dphi,\omega_s]$, the drift velocity $\ve V$ is given by
\begin{align}
\ve V&=\Big[\omega_s\,,\;-\frac{\Co}{\Io\st}\Big(\omega_s + \frac{\ApNew}{\AgNew}|\mathscr A|\sv^2\dphi\Big)\Big]\,,
\end{align}
and the stochastic driving takes the form $\delta\ve F=[0,\delta f]$. Here $\delta f$ is Gaussian white noise with vanishing mean and variance $\langle\delta f^2\rangle= 2D\delta t$, where $D$ is determined by the integral of the fluid-velocity correlations:
\begin{align}
D=\frac{\Co}{\Io\st}|\mathscr A|^2\ApNew{}^2\sv^2 \frac{1}{2}\int_{-\infty}^{\infty}\!\!\!\!{\rm d}t \,C_u(t)\,.
\end{align}
For platelets we must take $\ve X=[\dphi,\omega_s,\omega_p]$,
\begin{align}
\ve V&=\Big[\omega_s\,,\;-\frac{\Co}{\Io\st}\Big(\omega_s + \frac{\ApNew}{\AgNew}|\mathscr A|\sv^2\dphi\Big)+\frac{\omega_p^2}{\dphi}\new{\,,\;-}\frac{\Co}{\Io\st}\omega_p - \frac{\omega_p\omega_s}{\dphi}\Big]\,,
\end{align}
and $\delta\ve F=[0,\delta f_2,\delta f_3]$ with $\langle\delta f_i\delta f_j\rangle=2\delta_{ij}D\delta t$.
From the corresponding Fokker-Planck equations we obtain recursions for  the moments of components of $\ve X$ that can be solved for the moments of $\dphi$.
For columns we find that  $\dphi$ is Gaussian distributed with variance
\begin{align}
\label{eq:5result}
\langle\dphi^2\rangle=|\mathscr A|\AgNew\ApNew\frac{1}{2}\int_{-\infty}^{\infty}\!\!\!\!{\rm d}t\, C_u(t)\,.
\end{align}
In the statistical model the integral evaluates to $\sqrt{2\pi}\ell^3\AgNew/(30\,\sv)$. \new{It follows} that the \new{tilt-angle} variance is proportional to $\sv^{-1}$ in regime \raisebox{-0.2mm}{\ding{196}}, see Eq.~(25) in the main text.
For platelets we infer that $\dphi^2$ is exponentially distributed with mean equal to twice the variance of rods.
Note that this factor of $2$ (or $f_\Lambda$)
emerged also in regimes \raisebox{-0.2mm}{\ding{193}} to \raisebox{-0.2mm}{\ding{195}}. In regime  \raisebox{-0.2mm}{\ding{196}} this factor
is a consequence of the fact that the distributions of $\dphi$ have the same parameter for columns and platelets.

In summary, particle inertia results in a slow decay of the tilt-angle variance at large Stokes numbers (regime \raisebox{-0.2mm}{\ding{196}}),
$
 \langle\delta\varphi^2\rangle \propto \sv^{-1}$.
 The same caveat as for regime \raisebox{-0.2mm}{\ding{195}} applies:  the settling time $\taus$ must be larger than the viscous     time $a^2/\nu$, a necessary condition for the steady approximation for the inertial torque to hold.

\section{Analytical evaluation of the tilt-angle variance  in the statistical model}
\label{sec:eval}
\begin{figure}[ht]
\begin{overpic}[width=16cm,clip]{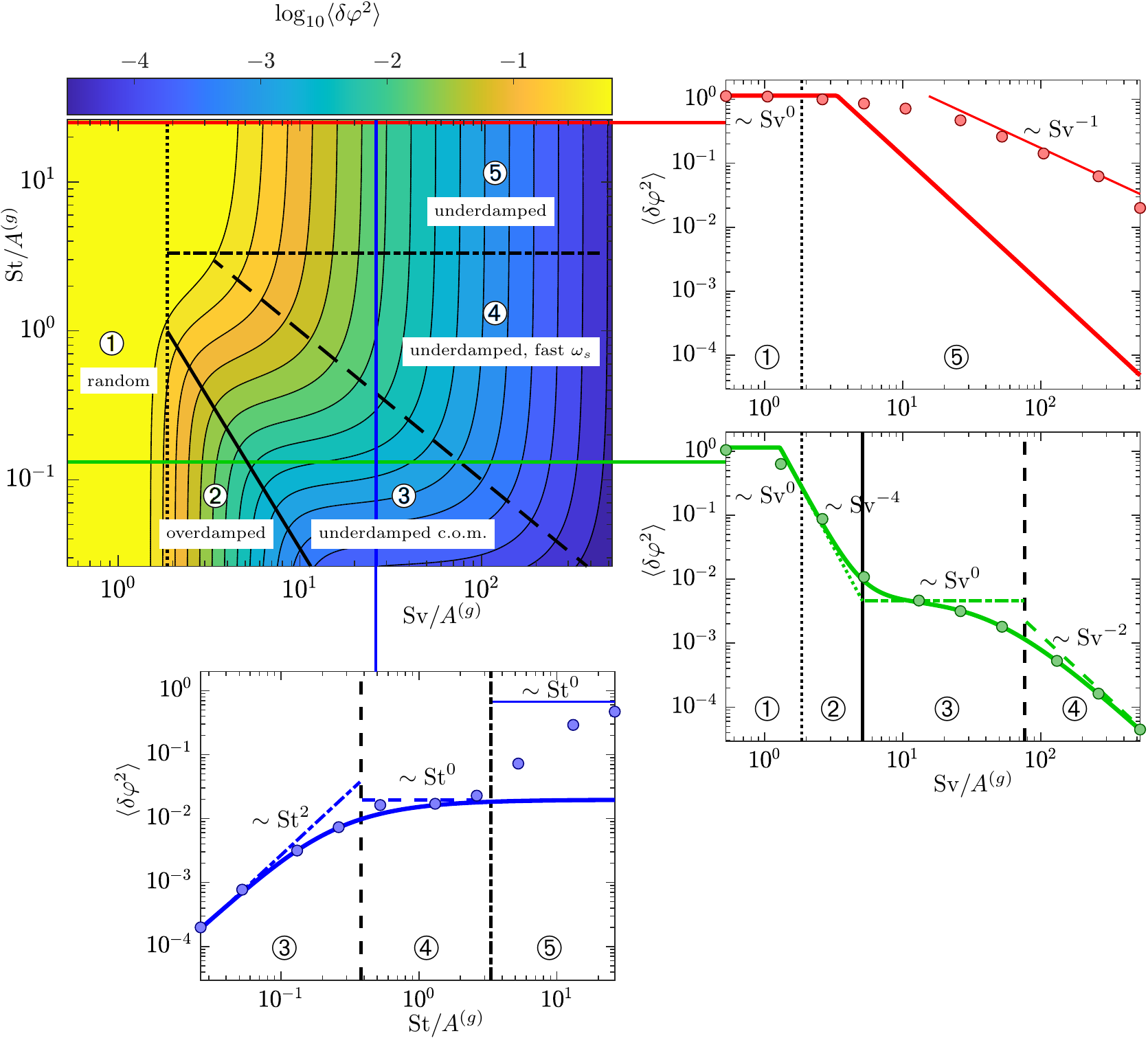}
\end{overpic}
\captionof{figure}{\label{fig:PhaseDiagramSM}
\new{
|Phase diagram of asymptotic regimes for the tilt-angle variance $\langle\dphi^2\rangle$ for regimes \raisebox{-0.2mm}{\ding{193}}--\raisebox{-0.2mm}{\ding{195}} based on \Eqnref{dphiSqrFinal}, evaluated with $C_0(t)$ at finite $\ku$, Eq.~(\ref{eq:C0Ku}). The variance $\langle\dphi^2\rangle$ is cut off at the value $\pi-2$ for uniformly distributed angles. Parameters and axis ranges are identical to Fig. 3 in the main text \new{(see text)}.
Subpanels show $\langle\dphi^2\rangle$ evaluated along slices with constant $\st/\AgNew$ (horizontal solid lines in the phase diagram) or constant $\sv/\AgNew$ (vertical solid line). The subpanels show data taken from Fig. 3 in the main text (markers), \Eqnref{dphiSqrFinal} (thick solid lines), and the asymptotes (24a) (dotted line), (24b) (dash-dotted lines), (24c) (dashed lines), and (25) (thin solid lines) from the main text.
}
}
\end{figure}
In the statistical model \citep{Gus16} the fluid-velocity field is generated from a vector potential, $\ve u(\ve x,t)=\ve\nabla\wedge\ve \Psi/\sqrt{6}$, where the components $\Psi_j(\ve x,t)$ of $\ve \Psi$ are Gaussian random functions with vanishing mean and correlation functions (in units made dimensionless using the Kolmogorov scales)
\begin{align}
\langle \Psi_i(\ve x,t)\Psi_j(\ve x,t)\rangle=\delta_{ij}\frac{\ell^4}{5}\exp\left[-\frac{|\ve x-\ve x'|^2}{2\ell^2}-\frac{|t-t'|}{\sqrt{5}\ku}\right]\,.
\end{align}
Differentiating this correlation function w.r.t. the spatial coordinate and evaluating the spatial coordinate along deterministic settling trajectories \eqnref{xxdDef} gives the following correlation functions needed to evaluate \Eqnref{phiSqrPlateletsUnevaluatedAppendix}
\begin{subequations}
\begin{align}
C_B(t)&=\frac{5+3\Lambda^2}{60}C_0(t)\Big[1-\frac{(1+|\Lambda|)(9+5|\Lambda|)}{2(5+3\Lambda^2)}\frac{t^2}{\taus^2}+\frac{(1+|\Lambda|)^2}{2(5+3\Lambda^2)}\frac{t^4}{\taus^4}\Big]\,,
\\
C_u(t)&=\frac{\ell^2}{15}C_0(t)\Big[1 - \frac{t^2}{2\taus^2}\Big]\,,
\eqnlab{uuCorr}
\\
C_X(t)&=\frac{\ell}{60}(5-3|\Lambda|) C_0(t) \frac{t}{\taus}\Big[1-\frac{1-|\Lambda|}{5-3|\Lambda|}\frac{t^2}{\taus^2}\Big]\,,\\
\mbox{with}&\quad C_0(t)=e^{-t^2/(2\taus^2) -|t|/(\sqrt{5}\ku)}\,.
\label{eq:C0Ku}
\end{align}
\end{subequations}
Inserting these correlation functions into \Eqnref{phiSqrPlateletsUnevaluatedAppendix} yields upon integration
\begin{align}
\begin{split}
\langle&\dphi^2\rangle=\frac{f_\Lambda}{30}\bigg[
\frac{2\AgNew{}^2 \st^2(5 + 3\Lambda^2) +\taus^2\AgNew{}^2 (\ApNew{}^2 - \AgNew{}^2) (1 - |\Lambda|)^2 +4(\ApNew{}^2 - \AgNew{}^2) \st^2 (1 + |\Lambda|)^2}{4 \ApNew{}^2 \mathscr A^2 \st^2 \sv^4}
\\
&+\taus^2 - \taus^2\frac{\AgNew(1 - |\Lambda|)}{|\mathscr A| \st \sv^2}
+\bigg((\AgNew{}^2 - \ApNew{}^2)\frac{\taus^4\AgNew{}^4 (1 - |\Lambda|)^2 +\taus^2\AgNew{}^2\st^2 (\Lambda^2 + 2|\Lambda|-3) +4\st^4 (1 + |\Lambda|)^2}{4 \taus^2\AgNew{}^2 \ApNew{}^2 \mathscr A^2 \st^2 \sv^4}
\\
&+ \frac{\taus^2\AgNew{}^2(1-|\Lambda|) - 2\st^2}{\AgNew |\mathscr A|\st \sv^2}
+\frac{\st^2}{\AgNew{}^2}-\taus^2
\bigg){\cal H}\bigg(\frac{\taus\AgNew}{\sqrt{2}\st}\bigg)
\bigg]
\end{split}\,,
\eqnlab{dphiSqrFinal}
\end{align}
where  ${\cal H}(x)=2x^2[1-\sqrt{\pi}x e^{x^2}{\rm erfc}(x)]$.
The result \eqnref{dphiSqrFinal} is valid for both platelets and columns  in regimes  \raisebox{-0.2mm}{\ding{193}} to  \raisebox{-0.2mm}{\ding{195}}.  It is written here using the limit $\ku\to\infty$, i.e. $C_0(t)=e^{-t^2/(2\taus^2)}$.
\new{\Figref{PhaseDiagramSM} shows the theory \eqnref{dphiSqrFinal} against $\sv/\AgNew$ and $\st/\AgNew$ with $\asp=0.1$, $\ell=10$, and $\ku=10$. \new{The theory is cut off at the tilt-angle variance of uniformly distributed angles, obtained by averaging $\dphi^2$ over the distribution, $\rho(\dphi)=|\sin\dphi|/2$ for $-\pi/2\le\dphi<\phi/2$. 

Comparing the analytical results plotted in \Figref{PhaseDiagramSM} to the statistical-model simulations in Fig.~3 (main text) shows} excellent agreement in regimes \raisebox{-0.2mm}{\ding{193}} to \raisebox{-0.2mm}{\ding{195}} where the theory applies. As expected, \Eqnref{dphiSqrFinal} fails in the limit of large $\st$. The upper inset in Figure~\ref{fig:PhaseDiagramSM} shows that the simulation results approach the asymptote~(25), derived from \Eqnref{5result} above, in the limit of large $\st$ and $\sv$ (regime \raisebox{-0.2mm}{\ding{196}}). The other insets confirm that the simulation results asymptotically approach the power-laws, Eqs.~(24).}

\section{Empirical parameters}
\begin{turnpage}
\begin{table}
\centering
\captionof{table}{\label{tab:parameters} Values of dimensionless parameters}
\begin{tabular}{llllllllllll}
\hline\hline
Reference &  $a$ & $\asp$ & $\varrho_{\rm p}$ & $\varrho_{\rm f}$ & $\nu$ &  $\mathscr{E}$  & ${\rm Fr}$ & $\st/\AgNew$ & $\sv/\AgNew$ & $\AgNew$ & $\rep$\cr
& [$\mu$m] & & [g/cm$^3$] & [g/cm$^3$] & [cm$^2$/s] & [cm$^2$/s$^3]$ & & & & \cr\\[-2.8mm]
\hline
\\[-2.8mm]
\cite{Bre04}          & 50 & $\tfrac{1}{6}$    &0.9&$1.2\times10^{-3}$&0.1&10   &0.01&0.08&7.9&0.86&0.4\cr
(hexagonal platelets) & 150 & $\tfrac{1}{12}$  &0.9&$1.2\times10^{-3}$&0.1&10   &0.01&0.37 &36 &0.85&5.4\cr
                      & 300 & $\tfrac{1}{16}$ &0.9&$1.2\times10^{-3}$&0.1&10   &0.01& 1.1 &$107 $&0.85&32\cr
                      & 50 & $\tfrac{1}{6}$    &0.9&$1.2\times10^{-3}$&0.1&1000 &0.32&0.81&2.5&0.86&0.4\cr
                      & 150 & $\tfrac{1}{12}$  &0.9&$1.2\times10^{-3}$&0.1&1000 &0.32&3.7 & 11.4 & 0.85&5.4\cr
                      & 300 & $\tfrac{1}{16}$ &0.9&$1.2\times10^{-3}$&0.1&1000 &0.32& 11  &34 &0.85&32\cr\cr
 \cite{Jucha2018} \& & 150 & $0.01$ & 0.92  & $1.4 \times 10^{-3}$ & 0.11 & 1 & 0.0018 & 0.01 & 5.9 & 0.85 & 0.46 \cr
 \cite{aurore} & 150 & $0.02$ & 0.92  & $1.4 \times 10^{-3}$ & 0.11 & 1 & 0.0018 & 0.02 & 11.8 & 0.85 & 0.93 \cr
 (disks)   & 150 & $0.01$ & 0.92  & $1.4 \times 10^{-3}$ & 0.11 & 16 & 0.014 & 0.04 & 3 & 0.85 & 0.46 \cr
  & 150 & $0.02$ & 0.92  & $1.4 \times 10^{-3}$ & 0.11 & 16 &0.014 & 0.08 & 5.9 & 0.85 & 0.93 \cr
  & 150 & $0.05$ & 0.92  & $1.4 \times 10^{-3}$ & 0.11 & 16 &0.014 & 0.2 & 14.8 & 0.85 & 2.3 \cr
 & 150 & $0.01$ & 0.92  & $1.4 \times 10^{-3}$ & 0.11 & 246 & 0.11 & 0.16 & 1.5 & 0.85 & 0.46 \cr
 & 150 & $0.02$ & 0.92  & $1.4 \times 10^{-3}$ & 0.11 & 246 & 0.11 & 0.33 & 3 & 0.85 & 0.93 \cr\cr
 \cite{Kramel} & $4.5\times 10^3$& $\tfrac{1}{20}$ &1.14 &0.995&$9.6\times 10^{-3}$&$0.4 \times 10^{-3}$ & $9\times 10^{-6}$ & $0.5\times 10^{-3}$ & 52 & 14.44 & 109 \cr
  (small triads)& $4.5\times 10^3$ & $\tfrac{1}{20}$ &1.14 &0.995&$9.6 \times 10^{-3}$ & $7.5 \times 10^{-3}$ & $83\times 10^{-6}$ & $2 \times 10^{-3}$ & 25 & 14.44 & 109 \cr
                       & $4.5\times 10^3$ & $\tfrac{1}{20}$ &1.14 &0.995&$9.6\times 10^{-3}$ & $9.5 \times 10^{-2}$ & $0.56 \times 10^{-3}$ & $7 \times 10^{-3}$ & 13 & 14.44& 109 \cr
                       & $4.5\times 10^3$ & $\tfrac{1}{20}$ &1.14 &0.995&$9.6\times 10^{-3}$ & 0.3 & $1.3\times 10^{-3}$ & $13 \times 10^{-3}$& 10& 14.44& 109 \cr
                       & $4.5\times 10^3$ & $\tfrac{1}{20}$ &1.14 &0.995&$9.6\times 10^{-3}$ & 0.5 & $1.9\times 10^{-3}$ & $17\times 10^{-3} $& 8.8& 14.44 & 109 \cr
                       & $4.5\times 10^3$ & $\tfrac{1}{20}$ &1.14 &0.995&$9.6\times 10^{-3}$ & 0.6 & $2.2\times 10^{-3}$ & $18.7 \times 10^{-3}$ & 8.4& 14.44& 109 \cr\cr
  \cite{esteban2020disks} &$5\times 10^{3}$&  $\tfrac{1}{20}$   & 2.7 & 0.996&0.01&14& 0.023& 0.36 & 15.4 & 0.85& 95 \cr
          (disks)                     &$7.5\times 10^{3}$& $\tfrac{1}{15}$ & 2.7 & 0.996& 0.01&14& 0.023 & 0.6 & 25.7 & 0.85 & 158\cr
\\[-2.8mm]\hline
\end{tabular}
\end{table}
\end{turnpage}

\cite{Bre04} considered alignment of hexagonal platelets of thickness $h$,
with diameters $d=2a$ in the range $0.1$ to $3\,$mm, and
aspect ratios $h/d \sim 2 d^{-0.55}$ with $h$ and $d$ in $\mu m$. In other words,
the aspect ratio varied from $\asp = 0.16$ for $d=100\,\mu$m  to $\asp = 0.024$ for $d=3\,$mm.
For these aspect ratios the shape coefficient $\AgNew$ is very close to the asymptotic
value at $\asp \rightarrow \infty$: $\AgNew \approx 0.85$.
Typical turbulent dissipation rates in clouds range from $\mathscr{E} \sim 1$ to $1000$ cm$^2$/s$^3$ \citep{grabowski1999comments}.
The mass density of air is  $\varrho_{\rm f} \sim 1.2$ kg/m$^3$,  with kinematic viscosity  $\nu \sim  10^{-5}$ m$^2$/s.
For a particle-mass density of  $\varrho_{\rm p}=0.9\times10^{-3}$ g/cm$^3$,  values of $\st$, $\sv$, $\rep$ are given in Table \ref{tab:parameters}, for $a=d/2=50,150,300\,\mu$m,
and for two values of  $\mathscr{E}$ ($\mathscr{E} =10$ cm$^2$/s$^3$ and  $\mathscr{E} =1000$ cm$^2$/s$^3$).

\cite{Jucha2018} and \cite{aurore}  analysed collisions of ice crystals settling in turbulence using model equations of motion in combination with DNS of turbulence, for
$\mathscr{E} \approx 1,16$, and $246\,$cm$^2$/s$^3$.

 \cite{Kramel} measured angular dynamics of triads settling in a turbulent water tank (fluid mass density $\varrho_{\rm f} =0.995$ g/cm$^3$, viscosity $\nu=0.96\times 10^{-6}$ m$^2$/s). The turbulent dissipation rate $\mathscr{E}$ ranged from $0.04$ to $60$ mm$^2$/s$^3$.
 The corresponding Kolmogorov time $\tauK$ ranges from
 $5\,$s to $0.13\,$s, and the Kolmogorov length $\etaK$ from  $2.16 \times 10^{-3}$ m  to $3.3\times 10^{-4}$ m. The small triads
 fabricated by Kramel were made out of slender fibres of radius $225 \,\mu$m and length $4.5\,$mm. A triad turns like a disk
 with radius   $4.5\,$mm and aspect ratio $1/20$, but with a different resistance coefficient  $\AgNew$ because the particle is ramified.
  We estimated the resistance coefficient $\AgNew$ from the experimentally observed settling speed in quiescent fluid using (in dimensional units)
$W = g\taup/\AgNew$.
The particles were only slightly heavier than the fluid, so we used  (in dimensional units)
\begin{align}
\label{eq:taup_nb}
\taup= \frac{2}{9} \Big( \frac{\varrho_{\rm p}}{\varrho_{\rm f}}-1\Big)\frac{a_\parallel a_\perp}{\nu}
\end{align}
instead of Eq.~(6) in the main text to estimate the particle response time.
With $W = 23.2\,$mm/s \citep{Kramel} we  found $\AgNew \approx 15.4$. We note that the other shape coefficients
too are quite different for a ramified particle compared with a solid platelet (Section \ref{sec:resistance}).
The particle Reynolds number is too large for the model to apply quantitatively, and we also note that the particles were larger than $\etaK$.

 \cite{esteban2020disks} studied the settling of small aluminium disks in a turbulent water tank ($\varrho_{\rm f} = 0.996\,$ g/cm$^3$, $\nu = 1.02\times 10^{-6} \,$m$^2$/s).
 Turbulent dissipation rate \citep{esteban2019laboratory} $\mathscr{E}=1.4 \times 10^{-3}\,$m$^2$/s$^3$; Kolmogorov scales
 $\tauK =0.026\,$s, $\etaK = 0.16\,$mm, $u_{\rm K} = 6.2\,$mm/s; Reynolds number ${\rm Re}_\lambda = 50$.
 The disks have diameter $d=10\,$mm, height $h=0.5\,$mm, as well as $d=15\,$mm, $h=1\,$mm.
 The particle-mass density is $\varrho_{\rm p} = 2.7\,$g/cm$^3$. Particle Reynolds numbers estimated from the Stokes settling velocity are very large, this shows that the small-$\rep$ approximation we used up to this point must break down. Instead we based our estimate in Table \ref{tab:parameters} upon the measured average settling speed. For the smaller disk, $\langle W \rangle = 9.5\,$cm/s, and for the larger disk $\langle W \rangle = 15.8\,$cm/s. Using this we estimate $\sv/\AgNew \sim  \langle W\rangle \tauK/\etaK = 15.4$ and $25.7$. The corresponding particle Reynolds numbers are   $95$ and $158$, and $\st /\AgNew\sim{\rm Fr}\,\sv/\AgNew = 0.36$ and $0.6$. While $\st /\AgNew$ and $\sv /\AgNew$ are appropriate measures of particle inertia and settling speed near the viscous limit, the regime we considered in this article, at large
 particle Reynolds numbers it is more appropriate to use the inertia parameter $I^\ast \propto \varrho_{\rm p} I_\perp/(\varrho_{\rm f}a^2)$, and the Archimedes number ${\rm Ar} = \sqrt{\tfrac{3}{16}(\varrho_{\rm p}/\varrho_{\rm f}-1)g a_\parallel}$\,\,\citep{auguste2013falling,Willmarth64,esteban2020disks}.

 \newpage
